\documentclass[aps,prl,twocolumn,superscriptaddress,showpacs,preprintnumbers,amsmath,amssymb]{revtex4-1}

\usepackage[tight,hang]{subfigure}
\usepackage{graphicx} 
\usepackage{dcolumn}  
\usepackage{comment}

\graphicspath{{ps}}

\newcommand{\bsg}{$b \rightarrow s \gamma$}
\newcommand{\BXsg}{$\overline{B} \rightarrow X_s \gamma$}
\newcommand{\BB}{$B \overline{B}$}
\newcommand{\qq}{$q \overline{q}$}
\newcommand{\upshiron}{$\Upsilon(4S)$}

\begin{document}

\preprint{\vbox{ \hbox{   }
						\vspace{2cm}
						\hbox{{\Large Belle Preprint 2014-15}\hspace{1cm}}   
                        \hbox{{\Large KEK Preprint 2014-26}\hspace{1cm}}  
}}
\begin{flushleft}
	\begin{figure}
	\hspace{-3cm}
		\includegraphics[width=3.7cm]{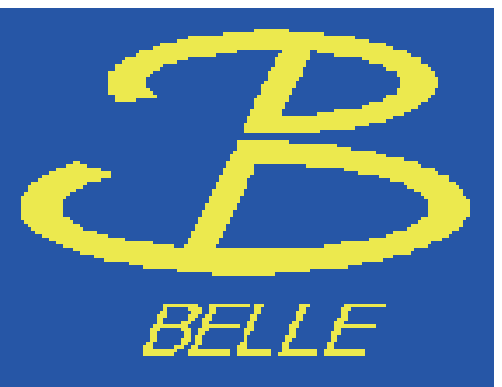}\label{fig:bellelogo}
	\end{figure}
\end{flushleft}

\vspace{-11cm}

\title{ \quad\\[1.0cm] Measurement of the {\BXsg} Branching Fraction with a Sum of Exclusive Decays}

\noaffiliation
\affiliation{University of the Basque Country UPV/EHU, 48080 Bilbao}
\affiliation{Beihang University, Beijing 100191}
\affiliation{University of Bonn, 53115 Bonn}
\affiliation{Budker Institute of Nuclear Physics SB RAS and Novosibirsk State University, Novosibirsk 630090}
\affiliation{Faculty of Mathematics and Physics, Charles University, 121 16 Prague}
\affiliation{University of Cincinnati, Cincinnati, Ohio 45221}
\affiliation{Deutsches Elektronen--Synchrotron, 22607 Hamburg}
\affiliation{Justus-Liebig-Universit\"at Gie\ss{}en, 35392 Gie\ss{}en}
\affiliation{The Graduate University for Advanced Studies, Hayama 240-0193}
\affiliation{Gyeongsang National University, Chinju 660-701}
\affiliation{Hanyang University, Seoul 133-791}
\affiliation{University of Hawaii, Honolulu, Hawaii 96822}
\affiliation{High Energy Accelerator Research Organization (KEK), Tsukuba 305-0801}
\affiliation{IKERBASQUE, Basque Foundation for Science, 48011 Bilbao}
\affiliation{Indian Institute of Technology Guwahati, Assam 781039}
\affiliation{Indian Institute of Technology Madras, Chennai 600036}
\affiliation{Indiana University, Bloomington, Indiana 47408}
\affiliation{Institute of High Energy Physics, Chinese Academy of Sciences, Beijing 100049}
\affiliation{Institute of High Energy Physics, Vienna 1050}
\affiliation{INFN - Sezione di Torino, 10125 Torino}
\affiliation{Institute for Theoretical and Experimental Physics, Moscow 117218}
\affiliation{J. Stefan Institute, 1000 Ljubljana}
\affiliation{Kanagawa University, Yokohama 221-8686}
\affiliation{Institut f\"ur Experimentelle Kernphysik, Karlsruher Institut f\"ur Technologie, 76131 Karlsruhe}
\affiliation{Kavli Institute for the Physics and Mathematics of the Universe (WPI), University of Tokyo, Kashiwa 277-8583}
\affiliation{Kennesaw State University, Kennesaw GA 30144}
\affiliation{Department of Physics, Faculty of Science, King Abdulaziz University, Jeddah 21589}
\affiliation{Korea Institute of Science and Technology Information, Daejeon 305-806}
\affiliation{Korea University, Seoul 136-713}
\affiliation{Kyungpook National University, Daegu 702-701}
\affiliation{\'Ecole Polytechnique F\'ed\'erale de Lausanne (EPFL), Lausanne 1015}
\affiliation{Faculty of Mathematics and Physics, University of Ljubljana, 1000 Ljubljana}
\affiliation{Luther College, Decorah, Iowa 52101}
\affiliation{University of Maribor, 2000 Maribor}
\affiliation{Max-Planck-Institut f\"ur Physik, 80805 M\"unchen}
\affiliation{School of Physics, University of Melbourne, Victoria 3010}
\affiliation{Moscow Physical Engineering Institute, Moscow 115409}
\affiliation{Moscow Institute of Physics and Technology, Moscow Region 141700}
\affiliation{Graduate School of Science, Nagoya University, Nagoya 464-8602}
\affiliation{Kobayashi-Maskawa Institute, Nagoya University, Nagoya 464-8602}
\affiliation{Nara Women's University, Nara 630-8506}
\affiliation{National Central University, Chung-li 32054}
\affiliation{National United University, Miao Li 36003}
\affiliation{Department of Physics, National Taiwan University, Taipei 10617}
\affiliation{H. Niewodniczanski Institute of Nuclear Physics, Krakow 31-342}
\affiliation{Niigata University, Niigata 950-2181}
\affiliation{Osaka City University, Osaka 558-8585}
\affiliation{Pacific Northwest National Laboratory, Richland, Washington 99352}
\affiliation{Peking University, Beijing 100871}
\affiliation{University of Pittsburgh, Pittsburgh, Pennsylvania 15260}
\affiliation{University of Science and Technology of China, Hefei 230026}
\affiliation{Seoul National University, Seoul 151-742}
\affiliation{Soongsil University, Seoul 156-743}
\affiliation{Sungkyunkwan University, Suwon 440-746}
\affiliation{School of Physics, University of Sydney, NSW 2006}
\affiliation{Department of Physics, Faculty of Science, University of Tabuk, Tabuk 71451}
\affiliation{Tata Institute of Fundamental Research, Mumbai 400005}
\affiliation{Excellence Cluster Universe, Technische Universit\"at M\"unchen, 85748 Garching}
\affiliation{Toho University, Funabashi 274-8510}
\affiliation{Tohoku University, Sendai 980-8578}
\affiliation{Department of Physics, University of Tokyo, Tokyo 113-0033}
\affiliation{Tokyo Institute of Technology, Tokyo 152-8550}
\affiliation{Tokyo Metropolitan University, Tokyo 192-0397}
\affiliation{University of Torino, 10124 Torino}
\affiliation{CNP, Virginia Polytechnic Institute and State University, Blacksburg, Virginia 24061}
\affiliation{Wayne State University, Detroit, Michigan 48202}
\affiliation{Yamagata University, Yamagata 990-8560}
\affiliation{Yonsei University, Seoul 120-749}
  \author{T.~Saito}\affiliation{Tohoku University, Sendai 980-8578} 
  \author{A.~Ishikawa}\affiliation{Tohoku University, Sendai 980-8578} 
  \author{H. Yamamoto}\affiliation{Tohoku University, Sendai 980-8578} 
  \author{A.~Abdesselam}\affiliation{Department of Physics, Faculty of Science, University of Tabuk, Tabuk 71451} 
  \author{I.~Adachi}\affiliation{High Energy Accelerator Research Organization (KEK), Tsukuba 305-0801}\affiliation{The Graduate University for Advanced Studies, Hayama 240-0193} 
  \author{H.~Aihara}\affiliation{Department of Physics, University of Tokyo, Tokyo 113-0033} 
  \author{S.~Al~Said}\affiliation{Department of Physics, Faculty of Science, University of Tabuk, Tabuk 71451}\affiliation{Department of Physics, Faculty of Science, King Abdulaziz University, Jeddah 21589} 
  \author{K.~Arinstein}\affiliation{Budker Institute of Nuclear Physics SB RAS and Novosibirsk State University, Novosibirsk 630090} 
  \author{D.~M.~Asner}\affiliation{Pacific Northwest National Laboratory, Richland, Washington 99352} 
 \author{T.~Aushev}\affiliation{Moscow Institute of Physics and Technology, Moscow Region 141700}\affiliation{Institute for Theoretical and Experimental Physics, Moscow 117218} 
% ...
\collaboration{The Belle Collaboration}


\begin{abstract}
We use 772$\times 10^6$ {\BB} meson pairs collected at the $\Upsilon(4S)$ resonance with the Belle detector to measure the branching fraction for {\BXsg}.
Our measurement uses a sum-of-exclusives approach in which 38 of the hadronic final states with strangeness equal to $+1$, denoted by $X_s$, are reconstructed. 
The inclusive branching fraction for $M_{X_s}<$\,2.8\,GeV/$c^2$, which corresponds to a minimum photon energy of 1.9 GeV, is measured to be ${\cal B}(${\BXsg}$)=(3.51\pm0.17\pm0.33)\times10^{-4}$, where the first uncertainty is statistical and the second is systematic.
\end{abstract}


\maketitle

\tighten

{\renewcommand{\thefootnote}{\fnsymbol{footnote}}}
\setcounter{footnote}{0}

\section{Introduction}
\label{sec:intro}
The {\bsg} transition is a flavor-changing neutral current process forbidden at tree level in the Standard Model (SM).
It proceeds at low rate through radiative penguin loop diagrams. 
Since the loop diagram is the only contribution, effects of new particles within the loop predicted by many new physics models (NP) can be investigated by a precise measurement of the branching fraction \cite{SMprediction1, NP1, NP2}. 
The inclusive branching fraction of the {\bsg} transition is sensitive to NP as it is theoretically well described in the SM. 
The SM calculation for the branching fraction has been performed at next-to-next-to-leading order in the perturbative expansion; for a photon energy above 1.6 GeV in the $B$ meson rest frame, the calculations predict a branching fraction of ${\cal B}\,(\overline{B} \rightarrow X_s \gamma) = (3.15 \pm 0.23) \times 10^{-4}$ \cite{SMprediction1} or $(2.98 \pm 0.26) \times 10^{-4}$ \cite{SMprediction2}, where $\overline{B}$ is either $\overline{B}^{0}$ or $B^-$ and $X_s$ denotes all the hadron combinations that carry strangeness of $+1$. 
Charge conjugation is implied throughout this article.
The dominant uncertainty of the expectation comes from the non-perturbative corrections.
The increase in background with decreasing photon energy limits the ability to make measurements below a minimum photon energy.
The measured branching fraction is extrapolated to a photon energy threshold of 1.6\,GeV to compare with the theoretical expectation. 
The current measured world average is ${\cal B}\,(\overline{B} \rightarrow X_s \gamma) = (3.40 \pm 0.21) \times 10^{-4}$ \cite{PDG}.
This value is consistent with the SM prediction within the uncertainties.

We report a measurement of the branching fraction of {\BXsg} with a 711\,fb$^{-1}$ data set collected at the {\upshiron}
resonance containing $772 \times 10^6$ {\BB} meson pairs recorded by the Belle detector \cite{Belle} at the KEKB asymmetric-energy
$e^+e^-$ collider (3.5\,GeV $e^+$ and 8.0\,GeV $e^-$) \cite{KEKB}. 
Our measurement uses a sum-of-exclusives approach whereby we measure as many exclusive final states of the $s$-quark hadronic system, $X_s$, as possible and then calculate the combined branching fraction. 
Exclusive branching fractions measured to date do not saturate the inclusive process, but one can still infer the total branching fraction by estimating the proportion of unmeasured modes using simulated fragmentation processes.

In this article, we present a measurement that supersedes an earlier Belle analysis \cite{ushiroda}, which was limited to only 5.8 fb$^{-1}$. We also use an improved analysis procedure here.

\section{Detector}
The Belle detector is a large-solid-angle magnetic spectrometer that consists of a silicon vertex detector (SVD),
a 50-layer central drift chamber (CDC), an array of aerogel threshold Cherenkov counters (ACC), a barrel-like arrangement of time-of-flight scintillation counters (TOF), and an electromagnetic calorimeter comprised of CsI(Tl) crystals (ECL) located inside a super-conducting solenoid coil that provides a 1.5\,T magnetic field.  
An iron flux-return located outside of the coil is instrumented to detect $K_L^0$ mesons and to identify muons (KLM).  
The detector is described in detail elsewhere \cite{Belle}.
The origin of the coordinate system is defined as the position of the nominal interaction point (IP). 
The $z$ axis is aligned with the direction opposite the $e^+$ beam and is parallel to the direction of the magnetic field within the solenoid. 
The $x$ axis is horizontal and points towards the outside of the storage ring; the $y$ axis is vertical upward. 
The polar angle $\theta$ and azimuthal angle $\phi$ are measured relative to the positive $z$ and $x$ axes, respectively.

\section{Simulation Sample}
We use Monte Carlo (MC) simulations to model signal and background events and to optimize the selection criteria prior to examining the signal region in the data.

We generate two types of signal MC samples, according to the $X_s$ mass region: one in the $K^*$(892) region ($M_{X_s} <$\,1.15\,GeV/$c^2$), where the hadronic part corresponds to a $K^*$, and the other in the inclusive hadronic region ($M_{X_s} >$\,1.15\,GeV/$c^2$). 
In the inclusive signal MC, various resonances and final states are simulated.
The photon energy spectrum is produced following the Kagan-Neubert model \cite{KNmodel} in the inclusive signal MC. 
This model has two parameters: the $b$ quark mass ($m_b$) and the Fermi-motion parameter of the $b$ quark inside the $B$ meson ($\mu_{\pi}^2$).
The nominal values of these parameters are determined from a best fit to the Belle inclusive photon energy spectrum \cite{belle_inclusive}: $m_b$\,=\,4.440 GeV/$c^2$ and  $\mu_{\pi}^2$\,=\,0.750 GeV$^2$.
In the inclusive $X_s$ mass region, the generated light quark pair is fragmented into final state hadrons in PYTHIA \cite{Pythia}. 
The signal reconstruction efficiency depends on the particle content in the final state; thus, it is important to determine the breakdown of the final state using data.
We assume the branching fraction to be the current measured world average in order to optimize the background rejection.

We simulate the background using {\qq} and {\BB} MC samples where we generate $e^+e^- \rightarrow q\overline{q}\,(q = u, d, s, c)$ (``continuum") and $e^+e^- \rightarrow \Upsilon(4S) \rightarrow B\overline{B}$, respectively.
In the latter case, we assume equal production of charged and neutral $B$ meson pairs.

\section{$B$ meson reconstruction and Background Suppression}
We reconstruct the $B$ candidate from a high energy photon and one of the 38 $X_s$ final states listed in Table \ref{tbl:finalstate}.
\begin{table}[b]
	\caption{ Reconstructed $X_s$ final states, where charge conjugated modes are implicitly included. A $K_S$($\eta$) is reconstructed via $\pi^+\pi^-$($\gamma\gamma$) final state. Final states with $\pi^+\pi^-\pi^0$ implicity include intermediate $\eta\rightarrow\pi^+\pi^-\pi^0$ decays (i.e., here the $\eta$ is not reconstructed).}
	\label{tbl:finalstate}
	\begin{tabular}
		{c|l|c|l}
		\hline \hline
		Mode ID & Final State & Mode ID & Final State \\
		\hline
		1  & $K^+\pi^- $                 & 20  & $K_S^0\pi^+\pi^0\pi^0$      \\
		2  & $K_S^0\pi^+$                & 21  & $K^{+}\pi^+\pi^-\pi^0\pi^0$   \\
		3  & $K^+\pi^0$                  & 22  & $K_S^0\pi^+\pi^-\pi^0\pi^0$ \\
		4  & $K_S^0\pi^0$                & 23  & $K^+\eta$      \\
		5  & $K^+\pi^+\pi^-$             & 24  & $K_S^0\eta$      \\
		6  & $K_S^0\pi^+\pi^-$           & 25  & $K^+\eta\pi^-$      \\
		7  & $K^+\pi^-\pi^0$             & 26  & $K_S^0\eta\pi^+$      \\
		8  & $K_S^0\pi^+\pi^0$           & 27  & $K^+\eta\pi^0$      \\
		9  & $K^{+}\pi^+\pi^-\pi^-$      & 28  & $K_S^0\eta\pi^0$      \\
		10 & $K_S^0\pi^+\pi^+\pi^-$      & 29  & $K^+\eta\pi^+\pi^-$      \\
		11 & $K^+\pi^+\pi^-\pi^0$        & 30  & $K_S^0\eta\pi^+\pi^-$      \\
		12 & $K_S^0\pi^+\pi^-\pi^0$      & 31  & $K^+\eta\pi^-\pi^0$      \\
		13 & $K^{+}\pi^+\pi^+\pi^-\pi^-$ & 32  & $K_S^0\eta\pi^+\pi^0$      \\
		14 & $K_S^0\pi^+\pi^+\pi^-\pi^-$ & 33  & $K^+K^+K^-$      \\
		15 & $K^+\pi^+\pi^-\pi^-\pi^0$   & 34  & $K^+K^-K_S^0$      \\
		16 & $K_S^0\pi^+\pi^+\pi^-\pi^0$ & 35  & $K^+K^+K^-\pi^-$      \\
		17 & $K^{+}\pi^0\pi^0$           & 36  & $K^+K^-K_S^0\pi^+$      \\
		18 & $K_S^0\pi^0\pi^0$           & 37  & $K^+K^+K^-\pi^0$      \\
		19 & $K^+\pi^-\pi^0\pi^0$        & 38  & $K^+K^-K_S^0\pi^0$      \\
		\hline \hline
	\end{tabular}
\end{table}

A high-energy photon generates an electromagnetic shower in the ECL and is detected as an isolated energy cluster not associated with charged particles. 
We take the photon candidate with an energy in the center-of-mass (CM) frame between 1.8 and 3.4\,GeV. 
The photon candidate is required to be within the acceptance of the barrel ECL, $33^{\circ}<\theta <\,132^{\circ}$.
It must satisfy $E_9/E_{25}\geq$\,0.95, which is the ratio of energy deposition within the $3\times3$ cells to that in the $5\times5$ cells centered on the maximum-energy ECL cell of the cluster. 
To reject candidates that arise from $\pi^0$ and $\eta$ decays, the photon candidate is paired with all other photons in the event with energy above 40\,MeV.
We reject the candidate based on a likelihood formed as a function of the invariant mass of the two-photon system and the laboratory energy of the partner photon ($\pi^0/\eta$-veto). 
Furthermore, photon candidates with a two-photon invariant mass between 117 and 153\,MeV/$c^2$ are rejected irrespective of the likelihood.

Charged particles must have a distance of closest approach to the IP within $\pm$5~cm along the $z$ axis and 0.5~cm in the transverse plane, and a laboratory momentum above 0.1\,GeV/$c$.
Flavor identification of $K^{\pm}$ and $\pi^{\pm}$ \cite{PID} is based on a likelihood formed with information from the specific ionization in the CDC, the flight time measured by the TOF, and the response of the ACC.

Candidate $K_S^0$ mesons are formed from $\pi^+\pi^-$ pairs by a multivariate analysis with a neural network technique \cite{NB}.
The neural network uses the distance between two helices in the $z$ direction, the flight length in the $x$-$y$ plane, the angle between the $K_S^0$ momentum and the vector joining the $K_S^0$ decay vertex to the IP, the shorter distance in the $x$-$y$ plane between the IP and two child helices, the angle between the pion momentum and the laboratory-frame direction in the $K_S^0$ rest frame, and the pion hit information in the SVD and CDC.
The selection efficiency and purity, evaluated with the MC sample, are 87\,\% and 94\,\%, respectively, over the entire momentum region. 
We also require that the di-pion invariant mass fall within 10\,MeV/$c^2$ of the nominal $K_S^0$ mass \cite{PDG}. 
We do not include $K_L^0$ mesons nor $K_S^0 \rightarrow \pi^0\pi^0$ decays.

Candidate $\pi^0$ mesons are reconstructed from pairs of photons with energy greater than 50\,MeV in the laboratory frame. 
We require a minimum $\pi^0$ momentum of 100\,MeV/$c$. 
The candidates must have an invariant mass between 125 and 145\,MeV/$c^2$, corresponding to $\pm$1.5$\sigma$ around the nominal $\pi^0$ mass, and the cosine of the angle in the laboratory frame between the two photons must be below 0.4.

Candidate $\eta$ mesons are reconstructed from pairs of photons with energy greater than 100\,MeV and an invariant mass between 515 and 570\,MeV/$c^2$, corresponding to $\pm$2.0$\sigma$ around the nominal $\eta$ mass.
We require the $\eta$ momentum to be above 200\,MeV/$c$. 
The photons from $\eta$ candidates must have a helicity angle ($\theta_{{\rm hel}}$) satisfying $\cos\theta_{\textrm{hel}} < 0.8$; $\theta_{{\rm hel}}$ is defined as the angle between the photon momentum and laboratory-frame direction in the $\eta$ rest frame. 
Although $\eta \rightarrow \pi^+\pi^-\pi^0$ decays are not explicitly reconstructed, they are included implicitly in the final states if the final state is included in Table \ref{tbl:finalstate}. 

The ``$K4\pi$" modes ($K^{+}\pi^+\pi^+\pi^-\pi^-$, $K_S^0\pi^+\pi^+\pi^-\pi^-$, $K^+\pi^+\pi^-\pi^-\pi^0$ and $K_S^0\pi^+\pi^+\pi^-\pi^0$), corresponding to  identification numbers (mode IDs) 13-16 in Table \ref{tbl:finalstate}, and ``$K2\pi^0$" modes ($K^{+}\pi^0\pi^0$, $K_S^0\pi^0\pi^0$, $K^+\pi^-\pi^0\pi^0$, $K_S^0\pi^+\pi^0\pi^0$, $K^{+}\pi^+\pi^-\pi^0\pi^0$, and $K_S^0\pi^+\pi^-\pi^0\pi^0$), corresponding to mode IDs 17-22, have substantial background. 
Therefore, the momentum of the leading $\pi$ in the $K4\pi$ mode and the leading $\pi^0$ in the $K2\pi^0$ mode is required to be above 400\,MeV/$c$, and the momentum of the subleading $\pi$ is required to be above 250\,MeV/$c$.

We do not include $\omega \rightarrow \pi^0\gamma$ modes ($K\omega$, $K\omega\pi$, $K\omega2\pi$) nor $\eta' \rightarrow \rho^0\gamma$ modes ($K\eta', K\eta'\pi, K\eta'2\pi$) as the yields are too small to make useful measurements.
The 38 measured final states cover 56\,\% of the total $X_s$ rate, according to the MC simulation.
Assuming the $K^0$ meson decays equally into $K^0_L$ and $K^0_S$, the proportion of our measured final states is 74\,\%.

We combine the photon candidate and the $X_s$ candidate to form the $B$ candidate. 
The latter is selected using two kinematic variables defined in the {\upshiron} rest frame: the beam-energy-constrained mass, $M_{\rm bc}=\sqrt{(E_{\rm beam}/c^2)^2-|\overrightarrow{p_{B}}/c|^{2}}$, and the energy difference, $\Delta E = E_{B}-E_{\rm beam}$, where $E_{\rm beam}$ is the beam energy and ($E_{B}$, $\overrightarrow{p_{B}}$) is the reconstructed four-momentum of the $B$ candidate.
The $B$ momentum vector $\overrightarrow{p_{B}}$ is calculated without using the magnitude of the photon momentum according to $\overrightarrow{p_{B}} = \overrightarrow{p_{X_{s}}} + \overrightarrow{p_{\gamma}}/|\overrightarrow{p_{\gamma}}| \times (E_{\rm beam} - E_{X_{s}})$, since the $X_s$ momentum and the beam energy are determined with substantially better precision than that of the photon candidate.
We require $M_{\rm bc}>$\,5.24\,GeV/$c^2$ and $-$0.15\,GeV$<\Delta E<$\,0.08\,GeV; the $\Delta E$ selection is tightened to $-$0.10\,GeV$<\Delta E<$0.05\,GeV for the final states with $2\pi^0$ and $\eta\pi^0$ (mode IDs of 17-22, 27-28 and 31-32) due to the larger background.

A large background still remains after signal reconstruction, dominated by three  categories.
The first is from events with $D$ meson decay, especially $B \rightarrow D^{(*)}\rho^+$.
These give rise to a peak in the signal region of $M_{\rm bc}$.
In order to suppress such background, a $D$ veto is applied for candidates with $M_{X_s} >$\,2.0\,GeV/$c^2$.
$D$ meson candidates of the major decay modes are reconstructed with combinations of particles used in the $X_s$ reconstruction. 
The event is rejected if any of the $D$ meson candidates falls in a veto window around the $D$ mass.
 We set the central value and the width of the veto window depending on the charge of the $D$ candidate and whether or not the $D$ candidate is reconstructed in a mode with a $\pi^0$ or $\eta$ meson.
Consequently, 90\,\% (23\%) of the signal (background) is kept. 

\begin{figure}[tb]
	\begin{center}
		\includegraphics[width=6.5cm]{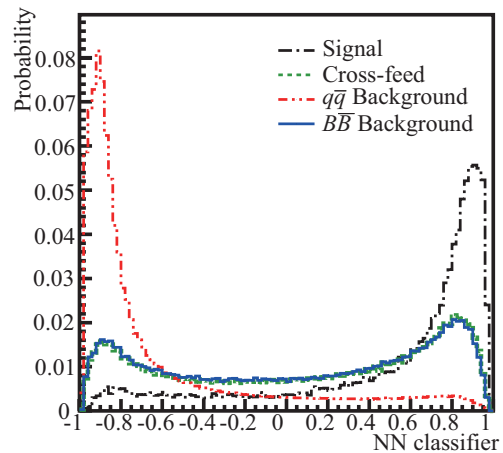}
		\caption[Neural Network output classifier (NN classifier) that ranges from $-1$ for the {\qq} background-like events to $+1$ for the signal-like events (MC). The signal (one dot-dashed line),  the cross-feed (dashed line), the {\qq} background (two dot-dashed line) and the {\BB} background (solid line)  are shown. We select events with the NN classifier above 0.78.]
		{\small Neural Network output classifier (NN classifier) that ranges from $-1$ for the {\qq} background-like events to $+1$ for the signal-like events (MC). The signal (one dot-dashed line),  the cross-feed (dashed line), the {\qq} background (two dot-dashed line) and the {\BB} background (solid line)  are shown. We select events with the NN classifier above 0.78.}
	    \label{fig:nboutput}
	\end{center}	
\end{figure}

The second and dominant category is continuum.
We reduce this background by applying a selection criterion based on the event shape. 
We perform a multivariate analysis with a neural network \cite{NB} that uses the following input variables: (1) the likelihood ratio of modified Fox-Wolfram moments \cite{FW,KSFW}, (2) the cosine of the angle between the $B$ candidate and the $z$ axis, (3) the angle between the thrust axis of the $B$ candidate's decay particles and that of the remaining particles in the event, (4) the thrust calculated with the remaining particles in the event, (5) the sphericity, (6) the aplanarity, (7) the flavor dilution factor of the accompanying $B$ meson that ranges from zero for no flavor information to unity for unambiguous flavor assignment \cite{flavortagging} and (8) the signal probability density for the $\Delta E$ value.
Variables (1)-(6) are calculated in the {\upshiron} rest frame.
The neural network is trained with signal and {\qq}-background MC events with  2.2\,GeV/$c^2<M_{X_s}<2.8$\,GeV/$c^2$.
The neural network output classifier (NN classifier) is also optimized in this region since the measurement here is difficult and incurs a large systematic uncertainty, while the signal in the low $M_{X_s}$ region is observed relatively easily. 
We obtain a NN classifier between $-1$ and $+1$, shown in Fig.\ref{fig:nboutput}, which achieves good separation between the signal and the {\qq} background.
The acceptance criterion of the NN classifier is optimized at 0.78 by the significance in 2.2 \,GeV/$c^2<M_{X_s}<2.8$\,GeV/$c^2$. 
After requiring that the NN classifier exceed this value, 52\,\% (2\,\%) of the signal ({\qq} background) is retained in the MC. 

The third major category of background is ``cross-feed'' from {\BXsg} events that have been incorrectly reconstructed. 
On average, there are approximately two $B$ meson candidates in a given event  after the {\qq} background suppression, since 38 final states are reconstructed concurrently.
To suppress cross-feed background, the $B$ candidate with the largest NN classifier for the {\qq} background suppression is selected (Best Candidate Selection or BCS); the efficiency is 85\,\% and the purity (defined as the ratio of number of signal to number of signal plus cross-feed after the BCS) is 68\,\%, evaluated with MC.

\section{Signal Yield Extraction}
To extract the signal yield, a fit is performed in the $M_{\rm bc}$ distribution with an unbinned maximum likelihood method. 
The likelihood function consists of PDFs for signal, cross-feed, peaking and non-peaking background from {\BB} events, and {\qq} background. 
The signal is modeled by a Crystal Ball function (CB) \cite{CB}:
\begin{equation}
	f_{\rm CB}(x) = \left \{
	\begin{array}{l}
		\exp\left(-\frac{1}{2}\left(\frac{x-m}{\sigma} \right)^2 \right) \ \ \ (\frac{x-m}{\sigma}>-\alpha) \\
		\frac{\left(\frac{n}{\alpha}\right)^n exp{({-\frac{1}{2}}\alpha^2})} {\left(\frac{n}{\alpha}-\alpha-\frac{x-m}{\sigma} \right)^n} \ \ \ (\frac{x-m}{\sigma}<-\alpha),
	\end{array}
	\right.
\end{equation}
where $m$ and $\sigma$ are the peak position and width, respectively, and the parameters $\alpha$ and $n$ characterize the non-Gaussian tail. 
These parameters are fixed to the values obtained from the large-statistics $B \rightarrow D\pi$ data sample, but a small correction to the tail parameters is applied by using the signal MC. 
For the cross-feed background, we construct a histogram PDF from the signal MC. 
The fraction of the cross-feed to the signal is fixed to the value obtained from the MC. 
A Gaussian function is adopted to model the peaking background. 
The shape parameters and yield are fixed to the values obtained from data within the $\pi^0/\eta$-veto window. 
The non-peaking background from {\BB} events is modeled by an ARGUS function  \cite{ARGUS}:
\begin{equation}
	\footnotesize
	f_{\rm ARG}(x) = x \biggl\{ 1-\left( \frac{x}{E_{\rm beam}^*}\right) ^2 \biggr\}^p \cdot  \exp\Biggl[ c \biggl\{ 1-\left( \frac{x}{E_{\rm beam}^*}\right) ^2\biggr\}\Biggr],
	\label{eq:argus}
\end{equation}
where the endpoint is fixed to the beam energy $E_{\rm beam}^*$ in the CM frame and other shape parameters and yield are allowed to float. 
For the {\qq} background PDF, we use a modified ARGUS function:
\begin{equation}
	\footnotesize
	f_{\rm ARG}^{\rm mod}(x) = x \biggl\{ 1-\left( \frac{x}{E_{\rm beam}^*}\right) ^q \biggr\}^p \cdot  \exp\Biggl[c \biggl\{1-\left(\frac{x}{E_{\rm beam}^*}\right)^2\biggr\}\Biggr],
\end{equation}
where a new power term, $q$, is introduced to account for the steep slope at low $M_{\rm bc}$. 
The shape and yield are determined via a fit to 90\,fb$^{-1}$ of off-resonance data collected at about 60\,MeV below the {\upshiron} resonance energy; the yield is scaled according to the luminosity. 
There are in total four free parameters in the fit for the signal extraction: the signal yield and the yield and shape parameters ($p$ and $c$) of the non-peaking background from {\BB} events.
We perform an ensemble test on toy MC to ensure no bias in the fitting procedure and verify with a full MC sample.
The signal yields are extracted in each $M_{X_s}$ bin.

\section{Calibration of $X_s$ Fragmentation Model}
Since the signal efficiency depends on the specific decay modes, the fragmentation model in the inclusive MC is calibrated to that of the data to reduce associated modeling systematic uncertainties. 
The final states are divided into ten categories, defined in Table \ref{tb:mode_category} \cite{Babar_semiincl}, to calibrate the MC selection efficiencies to those of data. 
\begin{table}[tb]
	\begin{center}
		  \caption[Mode category definitions for $X_s$ fragmentation check.]
		  {\small Mode category definitions for $X_s$ fragmentation check.}
		  \label{tb:mode_category}
          \begin{tabular}{ccc} \hline \hline
              Mode Category &  Definition   & Mode ID    \\ \hline 
              1    & $K\pi$ without $\pi^0$ &  1,2   \\ 
              2    & $K\pi$ with $\pi^0$    &  3,4   \\ 
              3    & $K2\pi$ without $\pi^0$&  5,6   \\ 
              4    & $K2\pi$ with $\pi^0$   &  7,8   \\ 
              5    & $K3\pi$ without $\pi^0$&  9,10   \\ 
              6    & $K3\pi$ with $\pi^0$   &  11,12   \\ 
              7    & $K4\pi$                &  13-16   \\ 
              8    & $K2\pi^0$               &  17-22   \\ 
              9    & $K\eta$                &  23-32   \\ 
              10   & 3$K$                   &  33-38   \\ \hline \hline
          \end{tabular}
	\end{center}
	\begin{center}
		  \caption[]
		  {\small The relative proportion (\%) of each mode in the range of 1.15\,GeV/$c^2<M_{X_s}<$\,2.8\,GeV/$c^2$ in the data, default MC and calibrated MC. Numbers in parentheses are deviation significances, defined as ([Proportion in MC]$-$[Proportion in data])/$\sigma_{\rm data}$. The uncertainties in the MC proportions are much smaller than those for the data proportions and can be neglected.}
		  \label{tb:fraction_data}
          \begin{tabular}{crrlrl} \hline \hline
              Mode  &     Data                  & \multicolumn{2}{c}{Default}  &  \multicolumn{2}{c}{Calibrated}  \\ 
			  Category&                            &  \multicolumn{2}{c}{MC}        &  \multicolumn{2}{c}{MC}  \\ \hline 
              1       &  4.2$\pm$0.4      & 10.3 &($+$17)       &    4.6 &($+$1.2) \\ 
              2       &  2.1$\pm$0.2      & 5.4 &($+$19)       &    2.4 &($+$1.6) \\ 
              3       &  14.5$\pm$0.5      & 12.9 &($-$3.1)      &    15.7 &($+$2.4) \\ 
              4       &  24.0$\pm$0.7      & 15.2 &($-$12)       &    24.0 &($-$0.0) \\ 
              5       &  8.3$\pm$0.8      & 5.9 &($-$3.3)      &    4.6 &($-$5.0) \\ 
              6       &  16.1$\pm$1.8      & 15.7 &($-$0.2)      &    19.2 &($+$1.8) \\ 
              7       &  11.1$\pm$2.8      & 12.3 &($+$0.4)      &    10.2 &($-$0.3) \\ 
              8       &  14.4$\pm$3.5      & 14.4 &($-$0.0)      &    11.6 &($-$0.8) \\ 
              9       &  3.2$\pm$0.8      & 4.9 &($+$2.3)      &    5.4 &($+$2.8) \\ 
              10      &  2.0$\pm$0.3      & 3.0 &($+$3.3)      &    2.3 &($+$1.0) \\ \hline \hline
          \end{tabular}
	\end{center}
\end{table}
In Table \ref{tb:fraction_data}, we compare the expected relative proportion of each category in data and MC. 
We find that the MC overestimates the fraction of the low multiplicity final-state $K\pi$ by more than a factor of two (i.e., mode categories 1 and 2).
In order to calibrate the proportions in the MC, we vary four relevant parameters in the nominal PYTHIA fragmentation model: the suppression factor of $s$-quark pair production  compared with $u$- or $d$-quark pair production (PARJ(2)), the probabilities for forming a spin-1 meson (PARJ(11), PARJ(15)), and an extra suppression factor for $\eta$ production in fragmentation (PARJ(25)). 
The proportions in the calibrated MC are given in the rightmost column of Table \ref{tb:fraction_data}. 
The total $\chi^2$ is improved from 826 to 47 using this calibration technique over the 10 degrees of freedom corresponding to the 10 decay mode categories. 
There are several decay-mode categories whose proportions in the MC deviate significantly from those in data, especially in category 5 ($K3\pi$ without $\pi^0$). 
We investigated the fragmentation proportions in four $M_{X_s}$ regions (1.15-1.5, 1.5-2.0, 2.0-2.4 and 2.4-2.8\,GeV/$c^2$) in the data and the calibrated MC shown in Table \ref{tb:fraction_data_MXs}.
However, several mode categories in the calibrated MC have large deviations from those in the data. 
\begin{table}[tb]
	\begin{center}
		  \caption[Relative proportions (\%) for each mode in each $M_{X_s}$ region in the data and calibrated MC. Numbers in parentheses are deviation significances, defined as (Proportion in MC$-$Proportion in data)/$\sigma_{\rm data}$.]
		  {\small Relative proportions (\%) for each mode in each $M_{X_s}$ region in the data and calibrated MC. Numbers in parentheses are deviation significances, defined as ([Proportion in MC]$-$[Proportion in data])/$\sigma_{\rm data}$.}
		  \label{tb:fraction_data_MXs}
		  \small
          \begin{tabular}{ccrl|crl} \hline \hline
                   & \multicolumn{3}{c|}{1.15$<M_{X_s}<$1.5\,(GeV/$c^2$)} &  \multicolumn{3}{c}{1.5$<M_{X_s}<$2.0\,(GeV/$c^2$)} \\ \hline
              Mode & Data         & \multicolumn{2}{c|}{MC} & Data & \multicolumn{2}{c}{MC} \\ \hline 
              1    & 10.7$\pm$0.6 & 14.6 &($+$6.4)  &  2.4$\pm$0.4 &  2.9& ($+$1.5)\\ 
              2    &  5.3$\pm$0.3 &  7.5 &($+$7.1)  &  1.2$\pm$0.2 &  1.5& ($+$1.7)\\ 
              3    & 25.7$\pm$0.8 & 21.6 &($-$5.0)  & 13.6$\pm$0.8 & 15.0& ($+$1.9)\\ 
              4    & 44.8$\pm$1.5 & 36.5 &($-$5.5)  & 19.7$\pm$1.1 & 22.0& ($+$2.2) \\ 
              5    &  0.9$\pm$0.5 &  1.0 &($+$0.1)  & 11.3$\pm$0.9 &  6.6& ($-$5.0) \\ 
              6    &  8.1$\pm$2.2 & 14.9 &($+$3.1)  & 21.7$\pm$2.4 & 23.7& ($+$0.9) \\ 
              7    &  0.3$\pm$0.5 &  0.5 &($+$0.5)  &  8.8$\pm$2.7 &  8.4& ($-$0.2) \\ 
              8    &  2.5$\pm$2.5 &  2.5 &($+$0.0)  & 14.7$\pm$2.1 & 12.2& ($-$1.2)\\ 
              9    &  1.7$\pm$0.4 &  0.9 &($-$1.8)  &  5.0$\pm$1.3 &  5.8& ($+$0.6) \\ 
              10   &  0.0$\pm$0.0 &  0.1 &($+$0.0)  &  1.6$\pm$0.2 &  1.3& ($-$1.5)\\ \hline
                   & \multicolumn{3}{c|}{2.0$<M_{X_s}<$2.4\,(GeV/$c^2$)} &   \multicolumn{3}{c}{2.4$<M_{X_s}<$2.8\,(GeV/$c^2$)} \\ \hline
              Mode & Data         & \multicolumn{2}{c|}{MC} & Data & \multicolumn{2}{c}{MC} \\ \hline 
              1    &  1.2$\pm$0.6 &  1.2& ($-$0.1)  &  0.5$\pm$0.7 &  0.9& ($+$0.7)\\ 
              2    &  0.6$\pm$0.3 &  0.6& ($+$0.0)  &  0.2$\pm$0.3 &  0.5& ($+$0.8) \\ 
              3    &  7.1$\pm$1.4 &  9.6& ($+$1.9)  &  3.8$\pm$2.2 &  8.2&($+$2.0)\\ 
              4    &  8.9$\pm$2.6 & 13.9& ($+$1.9)  &  8.5$\pm$4.0 & 11.8 &($+$0.8) \\ 
              5    & 12.1$\pm$2.5 &  8.3& ($-$1.5)  & 12.7$\pm$5.2 &  8.2& ($-$0.9) \\ 
              6    & 16.1$\pm$5.7 & 22.6& ($+$1.1)  &  3.3$\pm$12.8 & 21.2& ($+$1.4) \\ 
              7    & 28.0$\pm$9.1 & 16.5& ($-$1.3)  &  3.1$\pm$26.7 & 20.4& ($+$0.7) \\ 
              8    & 15.5$\pm$15.5& 18.5& ($+$0.2)  & 53.1$\pm$28.7 & 20.2& ($-$1.2)\\ 
              9    &  6.8$\pm$3.7 &  6.2& ($-$0.2)  & 10.6$\pm$8.2 & 5.9 &($-$0.6) \\ 
              10   &  3.6$\pm$1.1 &  1.4& ($-$2.0)  &  4.1$\pm$2.8 & 1.0& ($-$1.1)\\ \hline \hline            
          \end{tabular}
	\end{center}
\end{table}
\begin{table}
	\begin{center}
		  \caption[Calibration factors of each mode category in each $M_{X_s}$ region, which are ratios of the proportion in data to that in MC. ]
		  {\small Calibration factors of each mode category in each $M_{X_s}$ region, which are ratios of the proportion in data to that in MC.}
		  \label{tb:scale_directcalibration}
          \begin{tabular}{ccccc} \hline \hline
			  Mode     & \multicolumn{4}{c}{$M_{X_s}$ region\,(GeV/$c^2$)} \\
              Category &   1.15-1.5       & 1.5-2.0         & 2.0-2.4         & 2.4-2.8  \\ \hline 
              1        & 0.66$\pm$0.10    & 0.82$\pm$0.12   & 1.05$\pm$0.56   & 0.51$\pm$0.72 \\ 
              2        & 0.71$\pm$0.04    & 0.80$\pm$0.12   & 1.00$\pm$0.53   & 0.47$\pm$0.65 \\ 
              3        & 1.19$\pm$0.04    & 0.91$\pm$0.05   & 0.73$\pm$0.14   & 0.47$\pm$0.26 \\ 
              4        & 1.23$\pm$0.04    & 0.90$\pm$0.05   & 0.64$\pm$0.19   & 0.72$\pm$0.34 \\ 
              5        & 0.96$\pm$0.55    & 1.72$\pm$0.14   & 1.45$\pm$0.30   & 1.55$\pm$0.64 \\ 
              6        & 0.54$\pm$0.15    & 0.92$\pm$0.10   & 0.71$\pm$0.25   & 0.15$\pm$0.60 \\ 
              7        & 0.58$\pm$0.96    & 0.72$\pm$0.22   & 1.70$\pm$0.55   & 0.15$\pm$1.30 \\ 
              8        & 1.00$\pm$1.00    & 1.79$\pm$0.25   & 0.84$\pm$0.84   & 2.63$\pm$1.42 \\ 
              9        & 1.84$\pm$0.46    & 0.87$\pm$0.22   & 1.11$\pm$0.60   & 1.80$\pm$1.39 \\ 
              10       & 0.00$\pm$0.00    & 1.27$\pm$0.19   & 2.54$\pm$0.77   & 3.97$\pm$2.73 \\ \hline \hline
          \end{tabular}
	\end{center}
\end{table}
We find that the fine-tuning of the PYTHIA fragmentation is insufficient to accurately describe the data and, therefore, we calibrate directly using the ratio of the proportion for each mode category in data to that in the MC in the four mass regions shown in Table \ref{tb:scale_directcalibration}. 
The uncertainty on the ratio is the statistical uncertainty in fitting each mode category in data. 
Measurements of the $K2\pi^0$ signals (mode category 8) with 2.0\,GeV/$c^2< M_{X_s} <2.8$\,GeV/$c^2$ are difficult. 
Thus, the calibration by the factors in Table \ref{tb:scale_directcalibration} is not applied to mode category 8 in the range 2.0\,GeV/$c^2< M_{X_s} <2.8$\,GeV/$c^2$. 
The signal efficiencies in each $M_{X_s}$ bin before and after the calibration are reported in Table VI.
\begin{table}[tb]
	\begin{center}
		  \caption[The reconstruction efficiencies before and after the $X_s$ fragmentation calibration.]
		  {\small The reconstruction efficiencies before and after the $X_s$ fragmentation calibration.}
		  \label{tb:efficiency}
          \begin{tabular}{ccc} \hline \hline 
              $M_{X_s}$ bin  & Efficiency (\%)     & Efficiency (\%)  \\ 
              (GeV/$c^2$)    & before calibration & after calibration  \\ \hline 
              0.6-0.7        & 7.0                & 7.0   \\ 
              0.7-0.8        & 7.2                & 7.2   \\ 
              0.8-0.9        & 6.7                & 6.7   \\ 
              0.9-1.0        & 7.0                & 7.0   \\ 
              1.0-1.1        & 6.7                & 6.7   \\ 
              1.1-1.2        & 4.3                & 4.2   \\ 
              1.2-1.3        & 4.0                & 3.5   \\ 
              1.3-1.4        & 3.7                & 3.3   \\ 
              1.4-1.5        & 3.6                & 3.3   \\ 
              1.5-1.6        & 2.7                & 2.4   \\ 
              1.6-1.7        & 2.3                & 2.1   \\ 
              1.7-1.8        & 2.0                & 1.7   \\ 
              1.8-1.9        & 1.7                & 1.6   \\ 
              1.9-2.0        & 1.4                & 1.3   \\ 
              2.0-2.1        & 1.2                & 1.1   \\ 
              2.1-2.2        & 0.9                & 0.9   \\ 
              2.2-2.4        & 0.8                & 0.7   \\ 
              2.4-2.6        & 0.5                & 0.6   \\ 
              2.6-2.8        & 0.4                & 0.5   \\ \hline \hline
          \end{tabular}
	\end{center}
\end{table}

\section{Systematic Uncertainties}
The uncertainty on the total number of $B$ mesons in our data sample is 1.4\,\%.

The differences of the detector response between data and MC associated with photon detection, tracking of charged particles, $K_S^0$, $\pi^0$ and $\eta$ reconstruction, and $K^{\pm}/\pi^{\pm}$ identification are evaluated; the efficiencies are corrected by these values and the errors are taken as the systematic uncertainties on a bin-by-bin basis, as shown in Table \ref{tb:syst_MXs}.

The $D$-veto uncertainty is evaluated by using a control sample of $B \rightarrow X_sJ/\psi$ decays followed by $J/\psi \rightarrow {\ell^+}{\ell^-} ({\ell} = e, \mu)$. 
To mimic the conditions of a {\BXsg} decay, one of the leptons from the $J/\psi$ is combined with the $X_s$ candidate to ensure that the $X_s$ mass lies within the $D$-veto region.
This lepton is chosen such that the electric charge of $X_s$ is $-$1, 0 or 1 after the lepton is added; if the original $X_s$ is neutral, the lower-energy lepton is chosen. 
The invariant-mass distribution of the $D$ meson candidate in the $X_sJ/\psi$ control sample has a broad peak at the nominal $D$ mass region for $M_{X_s}>2.0$\,GeV/$c^2$, just as in the signal sample. 
The difference in the $D$-veto efficiency between the MC and data is evaluated by this control sample.

The uncertainty due to the {\qq} background suppression is evaluated using a control sample of $B \rightarrow D\pi$ decays, which provides sufficient statistical power ($\sim2\times10^{5}$ events). 
In the reconstruction of $B \rightarrow D\pi$, the prompt pion from the $B$ meson is treated as the photon candidate of the signal. 
The neural network is trained and optimized using the same methods as in the signal analysis. 
We take the difference in the efficiency between the data and MC as the systematic uncertainty.

The uncertainty due to the BCS is evaluated by again using a control sample of $B \rightarrow X_sJ/\psi\,(J/\psi \rightarrow {\ell^+}{\ell^-}, {\ell} = e, \mu)$ decays. 
The reconstruction procedure is the same as that in the $D$-veto uncertainty study. 
We assign the efficiency discrepancy between the MC and data as the systematic uncertainty.
The uncertainties of the $D$-veto, {\qq} background suppression and BCS are summed quadratically and given as ``Background rejection" in Table \ref{tb:syst_MXs}.

In order to evaluate the uncertainty due to the signal PDF, we use the variation in the signal yield after changing the parameter values by one standard deviation (1$\sigma$) around the central value of the fit. 
To evaluate the uncertainty from the cross-feed histogram PDF, we perform two sets of 1$\sigma$ systematic variations: on the bin-by-bin content in the PDF and on the ratio of the cross-feed yield to the signal yield.
The variation on the signal yield is taken as the systematic uncertainty.
To evaluate the uncertainty due to the peaking background PDF, we vary the parameter values by 1$\sigma$ and repeat the fits to data. 
To evaluate the uncertainty due to the {\qq} background PDF, we use the variation in the signal yield when varying the parameter values by 1$\sigma$ in the fits to the off-resonance data. 
The uncertainties due to the $M_{\rm bc}$ PDF are reported in Table \ref{tb:syst_MXs}.

The uncertainty due to the fragmentation model is determined by varying the decay channel proportions given in Table \ref{tb:fraction_data_MXs} by their respective uncertainties.
The exception is for the proportion of $K2\pi^0$ (mode category 8) in 2.0\,GeV/$c^2 <M_{X_s}< 2.8$\,GeV/$c^2$, where we use the proportions in MC and a variation of $^{+100\,\%}_{-50\,\%}$ as the uncertainty.
The fragmentation uncertainties for each $M_{X_s}$ bin are obtained by summing in quadrature the changes for each of the ten mode categories.
Since the threshold between $K^*$ and the inclusive $X_s$ used in the MC modeling is fixed at 1.15\,GeV/$c^2$, we change this boundary to 1.10\,GeV/$c^2$ and 1.20\,GeV/$c^2$ to evaluate the uncertainty due to the threshold. 
These uncertainties are included in the fragmentation uncertainty.

The proportion of missing final states that are not included in our reconstructed modes affects the reconstruction efficiency. 
The uncertainty on the relative proportion of each of the 38 measured final states is evaluated by varying the parameters of the fragmentation model used in the calibration of the MC within their allowed ranges as determined from data.
We take the difference from the nominal value as the systematic uncertainty on the missing proportion, as given in Table \ref{tb:syst_MXs}.

The following uncertainties are considered correlated across $M_{X_s}$ bins:
 the {\BB}-counting, detector-response and background-rejection uncertainties. 
We take the uncertainties on all $M_{\rm bc}$ PDFs except for the cross-feed to be uncorrelated across all mass bins, and the uncertainty on the cross-feed PDF is considered completely correlated. 
The total uncertainties due to the fragmentation and missing proportion factors are evaluated in four different $M_{X_s}$ mass regions (1.15-1.5, 1.5-2.0, 2.0-2.4, 2.4-2.8\,GeV/$c^2$). 
The uncertainties across these mass regions are considered to be uncorrelated, but the uncertainties across the mass bins within a given mass region are considered to be correlated. 

\begin{table*}[tb]
	\begin{center}
		  \caption[Systematic uncertainties\,(\%) in each $M_{X_s}$ mass bin.]
		  {\small Systematic uncertainties\,(\%) in each $M_{X_s}$ mass bin.}
		  \label{tb:syst_MXs}
		  \small
          \begin{tabular}{crrrrrrrrrr} \hline \hline
              $M_{X_s}$ bin &{\BB}\hspace{0.4cm}    &Detector & Background& Signal & Cross-feed& Peaking& {\qq} BG& Frag.& Missing\hspace{0.3cm}   &Total \\ 
              (GeV/$c^2$)   & counting&response & rejection & PDF    & PDF    &  BG PDF   &  PDF    &      & proportion  &    \\ \hline         
              0.6-0.7       &   1.4   & 2.7     &  3.4      &  0.0   &  0.0   &  0.0   &  0.0   & -       &  -         &  4.5   \\ 
              0.7-0.8       &   1.4   & 2.6     &  3.4      &  0.1   &  12.2  &  7.8   &  0.0   & -       &  -         &  15.3   \\ 
              0.8-0.9       &   1.4   & 2.6     &  3.4      &  0.2   &  0.4   &  0.5   &  0.0   & -       &  -         &  4.5   \\ 
              0.9-1.0       &   1.4   & 2.6     &  3.4      &  0.1   &  0.5   &  0.4   &  0.0   & -       &  -         &  4.5   \\ 
              1.0-1.1       &   1.4   & 2.6     &  3.4      &  0.1   &  2.9   &  1.1   &  0.3   & -       &  -         &  5.4   \\          
              1.1-1.2       &   1.4   & 3.0     &  3.4      &  0.4   &  3.1   &  1.7   &  0.2   &  32.1   &  1.2       &  32.1   \\         
              1.2-1.3       &   1.4   & 3.2     &  3.4      &  0.2   &  1.6   &  0.9   &  0.0   &  2.1    &  1.0       &  5.6   \\ 
              1.3-1.4       &   1.4   & 3.2     &  3.4      &  0.2   &  1.6   &  0.2   &  0.0   &  2.6    &  1.9       &  6.0  \\ 
              1.4-1.5       &   1.4   & 3.1     &  3.4      &  0.2   &  2.0   &  0.1   &  0.0   &  4.0    &  1.3       &  6.7  \\ 
              1.5-1.6       &   1.4   & 3.3     &  3.4      &  0.6   &  2.2   &  0.1   &  0.0   &  2.4    &  1.3       &  6.1   \\ 
              1.6-1.7       &   1.4   & 3.5     &  3.4      &  0.1   &  1.7   &  2.1   &  0.2   &  2.8    &  1.9       &  6.7   \\ 
              1.7-1.8       &   1.4   & 3.6     &  3.4      &  0.1   &  2.2   &  1.7   &  0.2   &  3.4    &  1.0       &  6.8   \\  
              1.8-1.9       &   1.4   & 3.7     &  3.4      &  0.1   &  1.9   &  2.0   &  0.1   &  3.6    &  2.1       &  7.2   \\ 
              1.9-2.0       &   1.4   & 3.7     &  3.4      &  0.1   &  4.2   &  4.0   &  0.1   &  3.7    &  1.6       &  8.8   \\  
              2.0-2.1       &   1.4   & 3.8     &  3.4      &  0.1   &  5.6   &  0.6   &  0.2   &  17.8   &  2.2       &  19.5   \\ 
              2.1-2.2       &   1.4   & 3.8     &  3.4      &  0.3   &  3.7   &  2.5   &  0.4   &  21.9   &  1.9       &  23.1   \\   
              2.2-2.4       &   1.4   & 3.8     &  3.4      &  0.1   &  7.4   &  7.1   &  0.0   &  25.5   &  1.6       &  28.0   \\ 
              2.4-2.6       &   1.4   & 3.8     &  3.4      &  0.1   &  11.5  &  21.8  &  0.3   &  29.6   &  1.0       &  38.9   \\ 
              2.6-2.8       &   1.4   & 3.8     &  3.4      &  0.1   &  44.7  &  101.0 &  0.9   &  29.4   &  2.0       &  113.9   \\ \hline \hline            
          \end{tabular}
	\end{center}
\end{table*}

\section{Branching Fractions}
\begin{figure*}[tb]
	\begin{center} 
		\subfigure[0.6\,$<M_{X_s}<$\,0.7\,(GeV/$c^2$)] {\includegraphics[width=7.0cm]{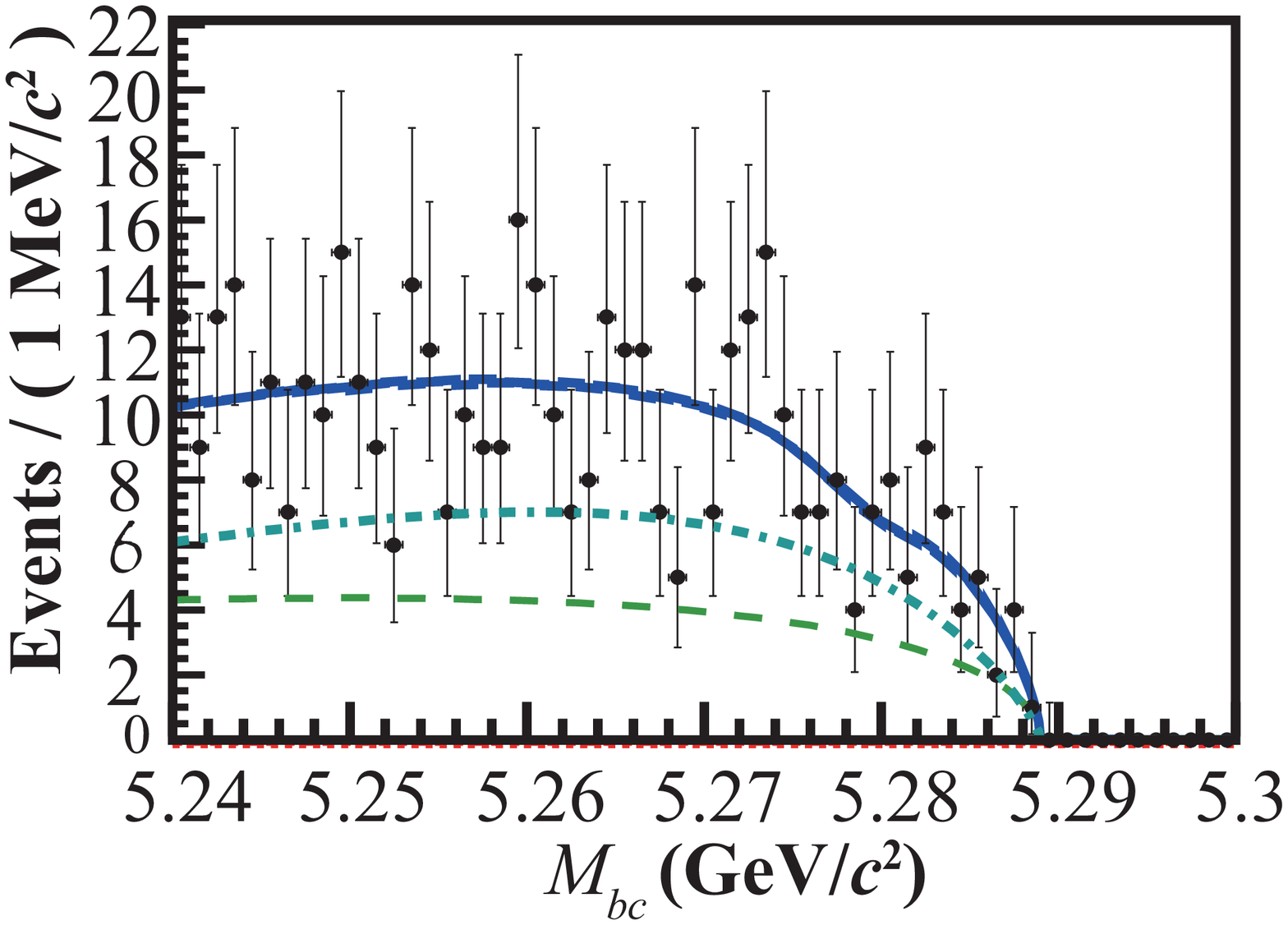} \label{fig:Mbcfit_MXs0.6-0.7}}
		\subfigure[0.7\,$<M_{X_s}<$\,0.8\,(GeV/$c^2$)] {\includegraphics[width=7.0cm]{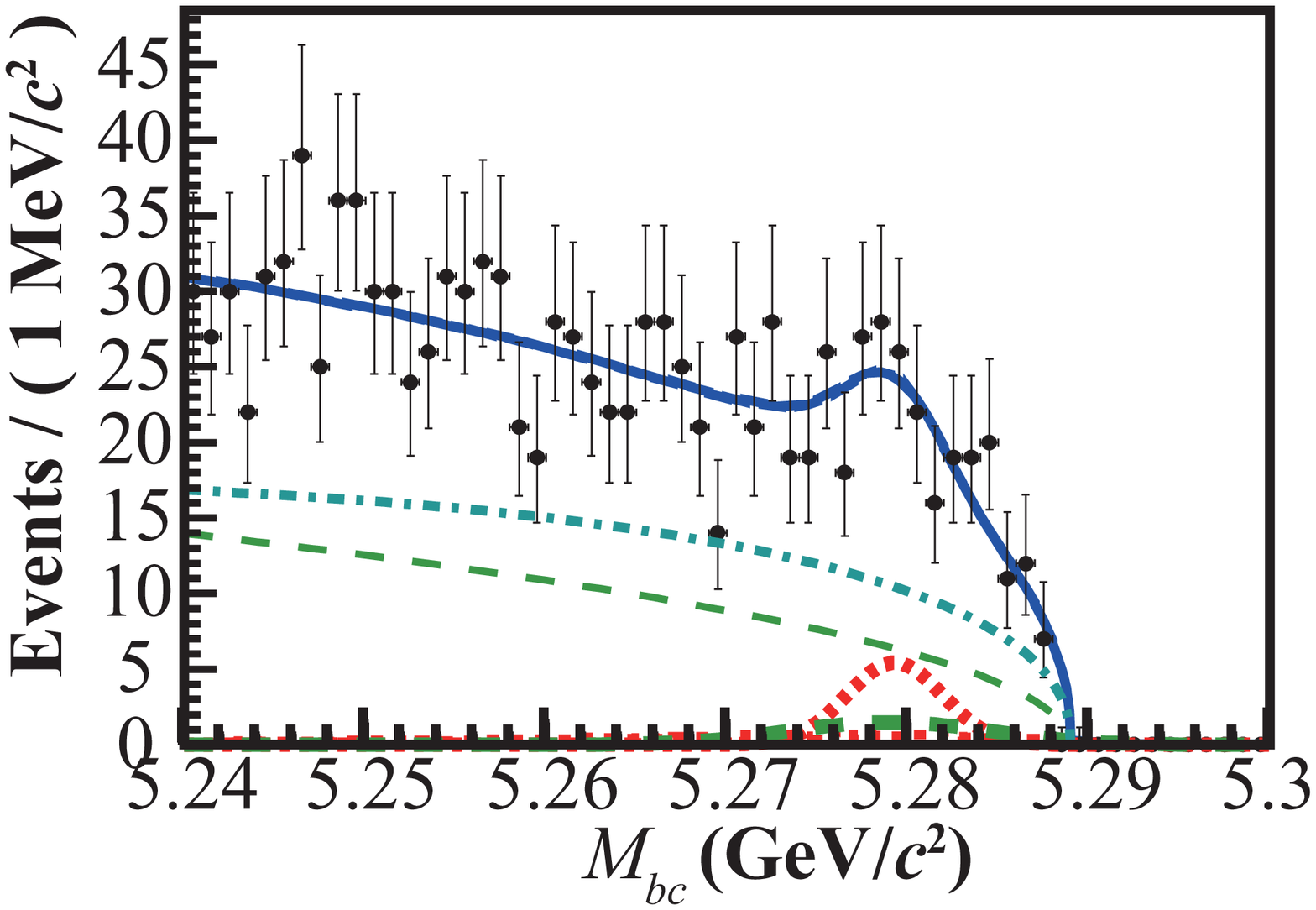} \label{fig:Mbcfit_MXs0.7-0.8}}
		\subfigure[0.8\,$<M_{X_s}<$\,0.9\,(GeV/$c^2$)] {\includegraphics[width=7.0cm]{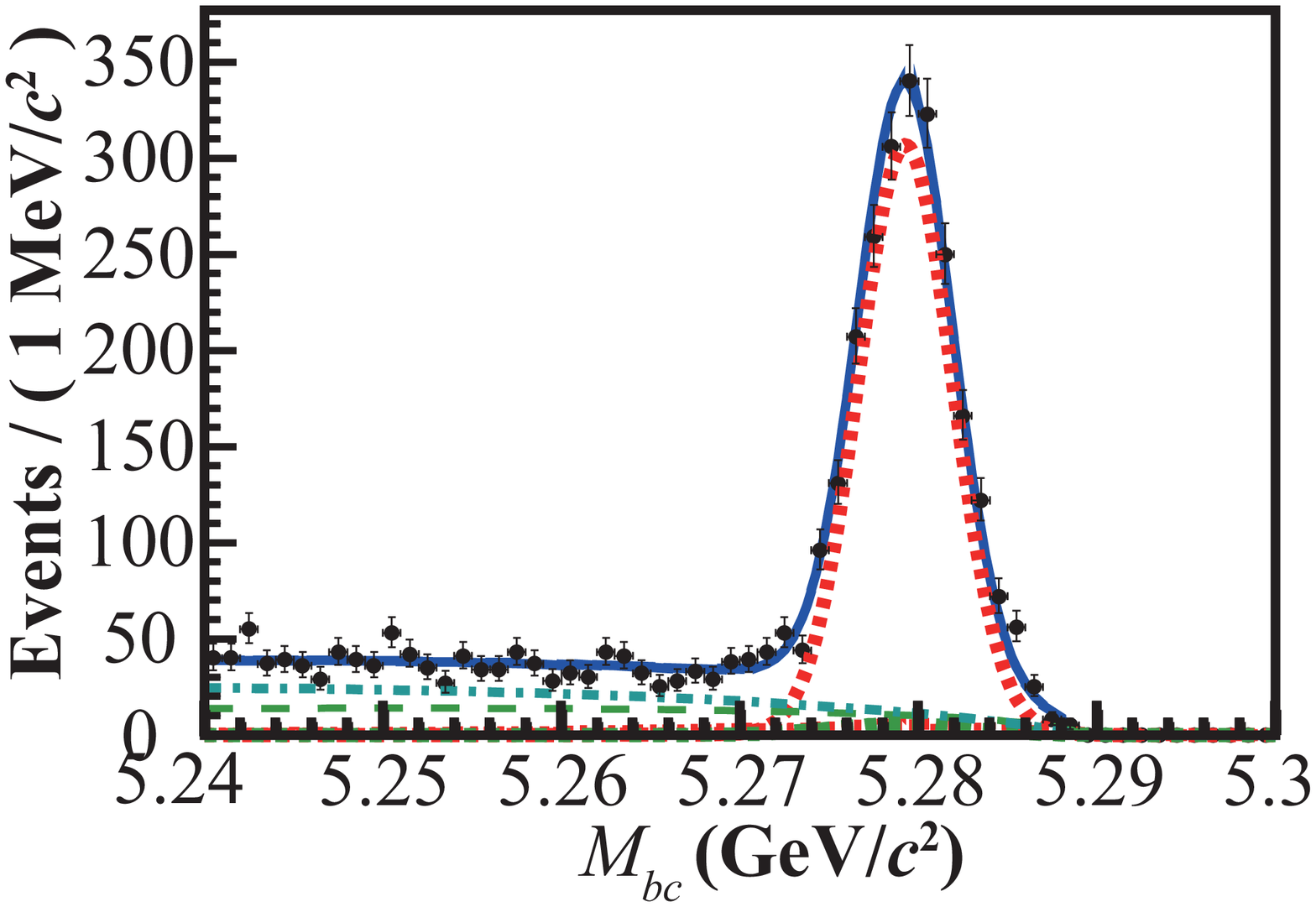} \label{fig:Mbcfit_MXs0.8-0.9}}
	    \subfigure[0.9\,$<M_{X_s}<$\,1.0\,(GeV/$c^2$)] {\includegraphics[width=7.0cm]{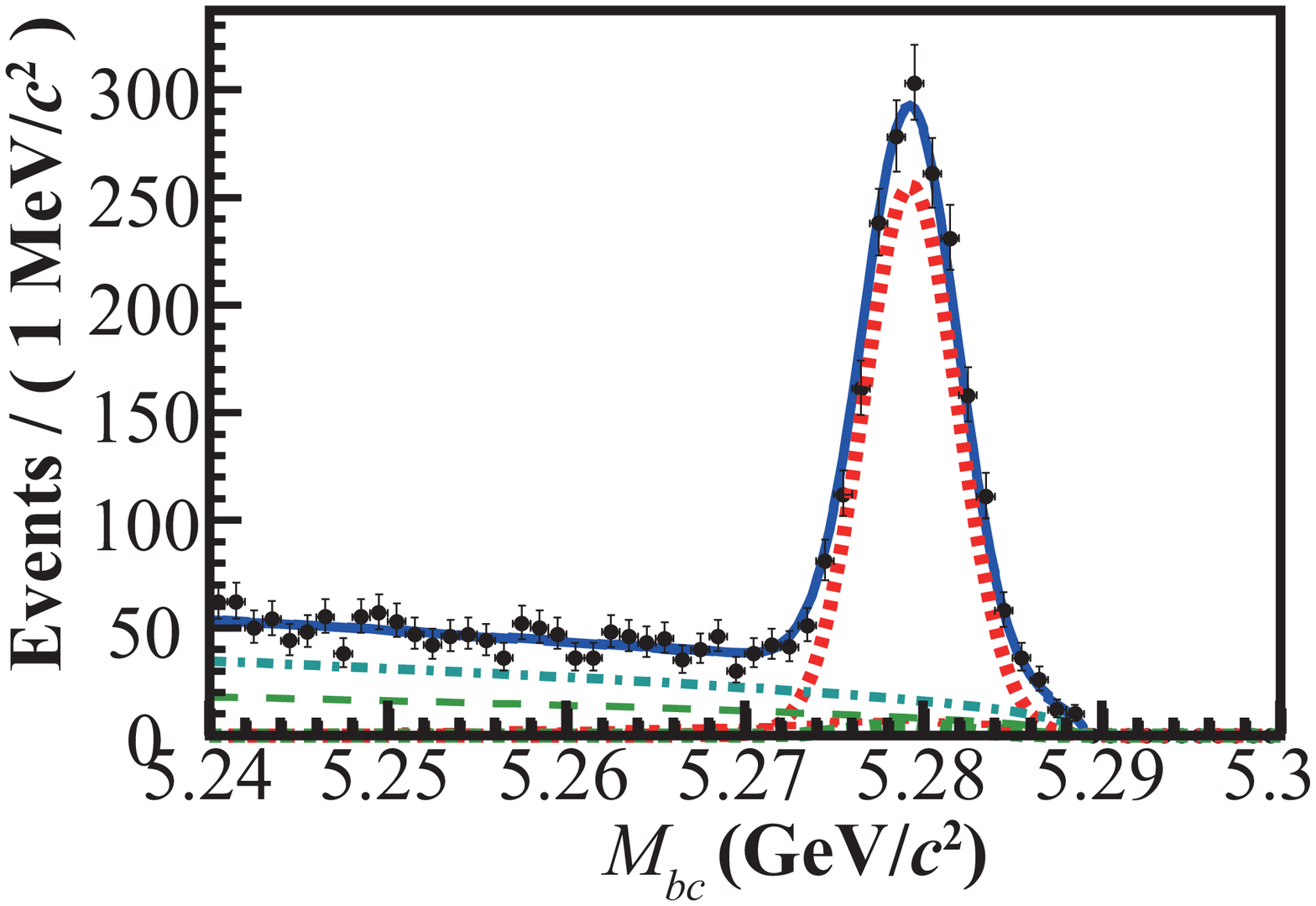} \label{fig:Mbcfit_MXs0.9-1.0}}
	    \subfigure[1.0\,$<M_{X_s}<$\,1.1\,(GeV/$c^2$)] {\includegraphics[width=7.0cm]{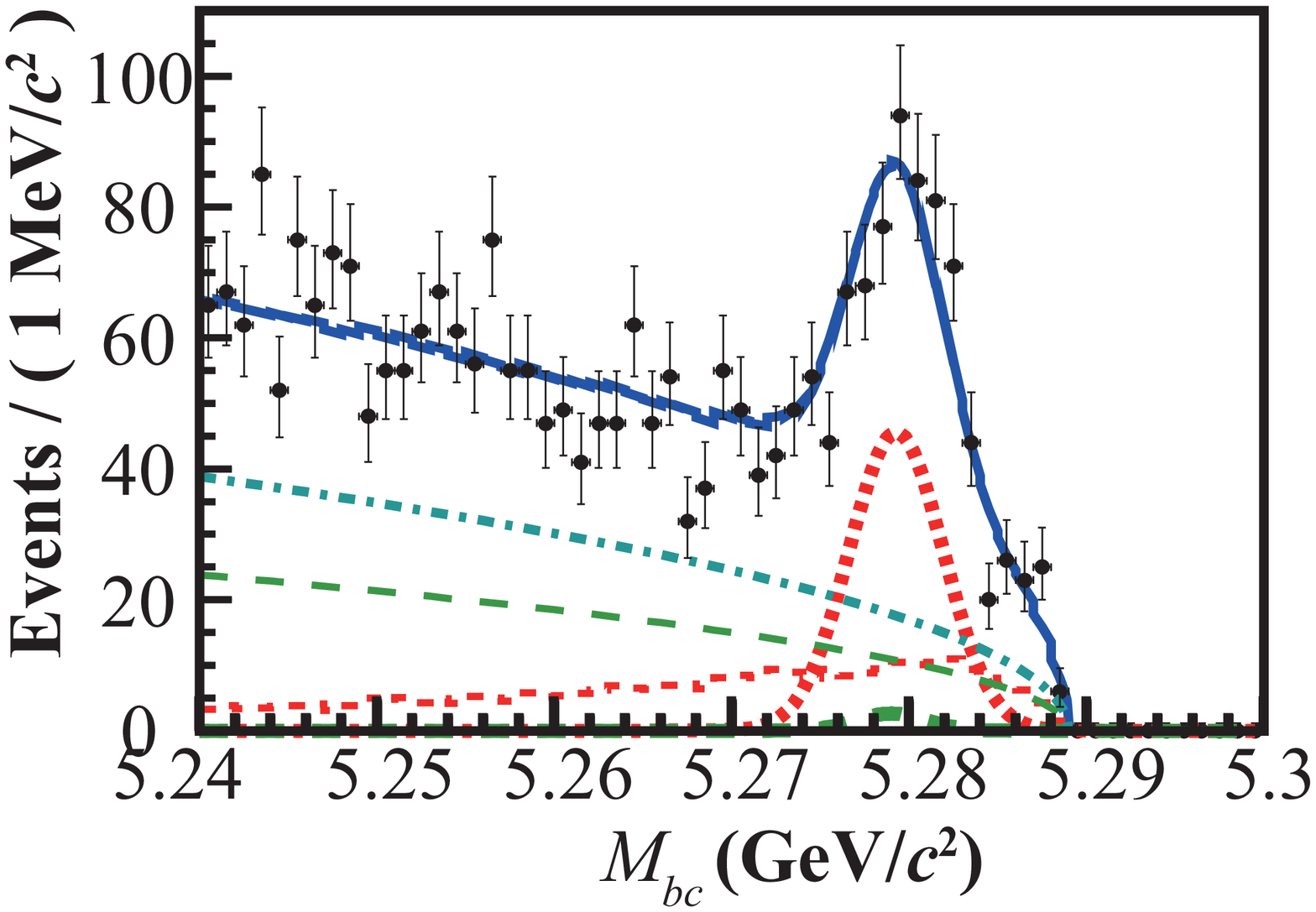} \label{fig:Mbcfit_MXs1.0-1.1}}
	    \subfigure[1.1\,$<M_{X_s}<$\,1.2\,(GeV/$c^2$)] {\includegraphics[width=7.0cm]{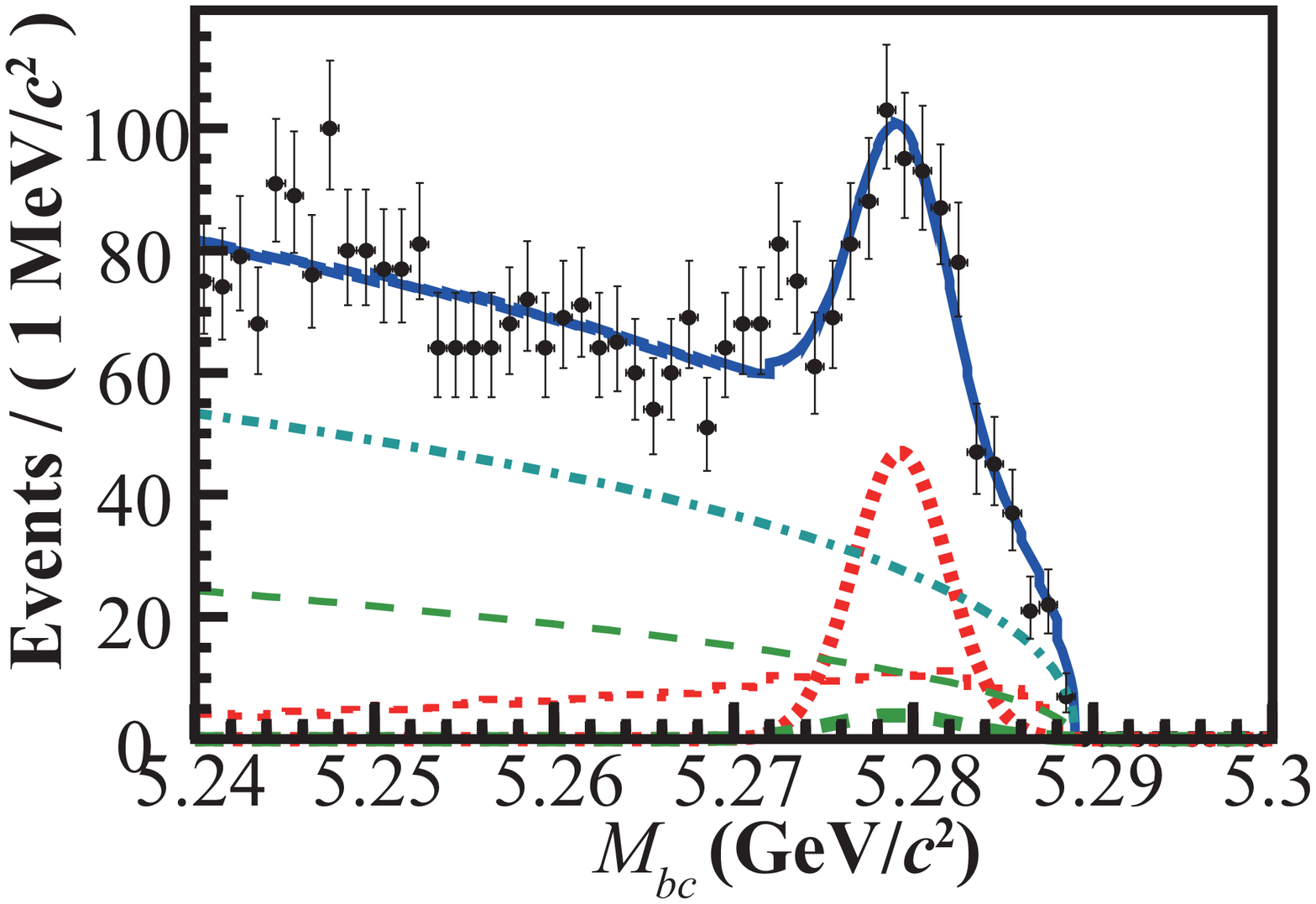} \label{fig:Mbcfit_MXs1.1-1.2}}
	    \subfigure[1.2\,$<M_{X_s}<$\,1.3\,(GeV/$c^2$)] {\includegraphics[width=7.0cm]{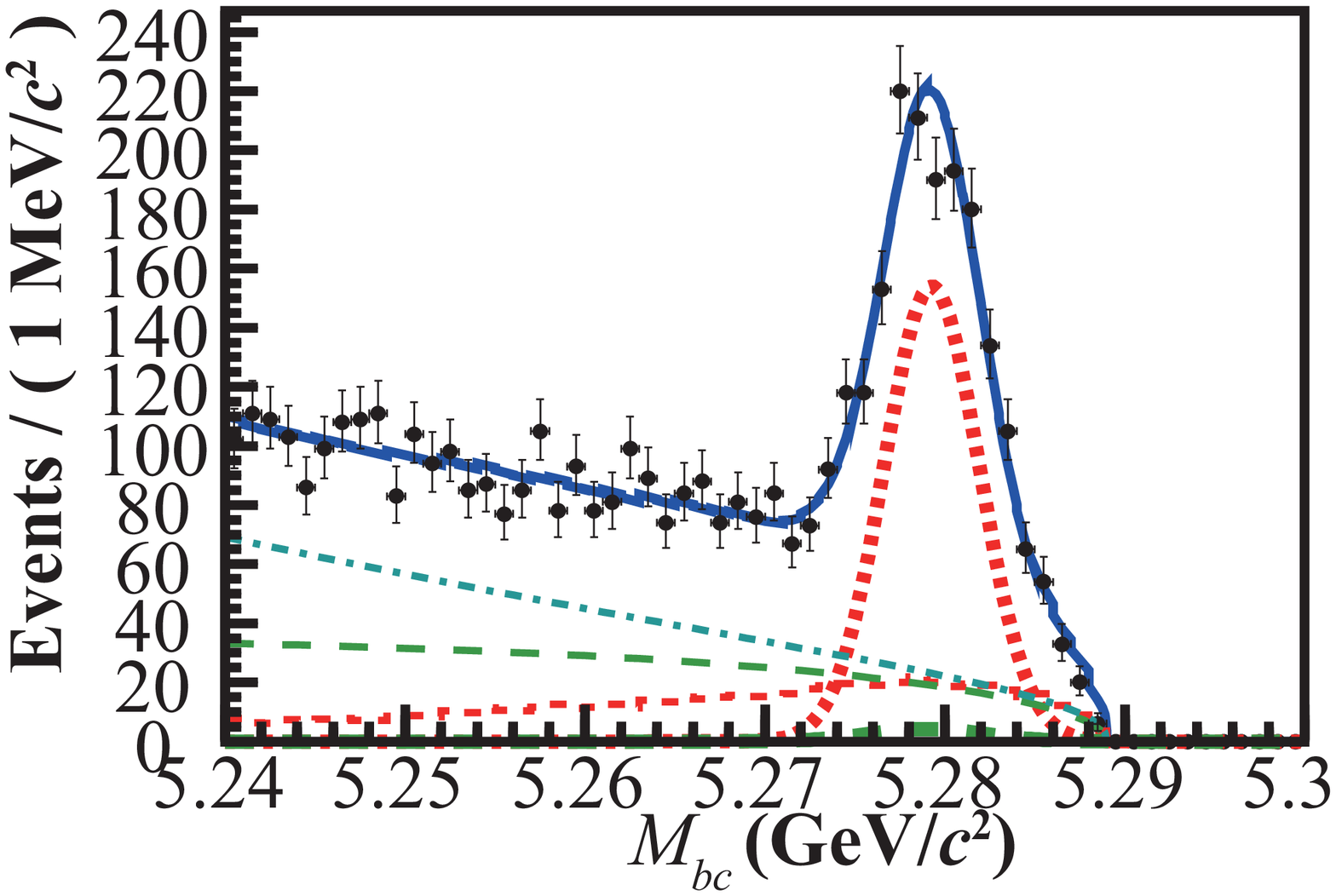} \label{fig:Mbcfit_MXs1.2-1.3}}
	    \subfigure[1.3\,$<M_{X_s}<$\,1.4\,(GeV/$c^2$)] {\includegraphics[width=7.0cm]{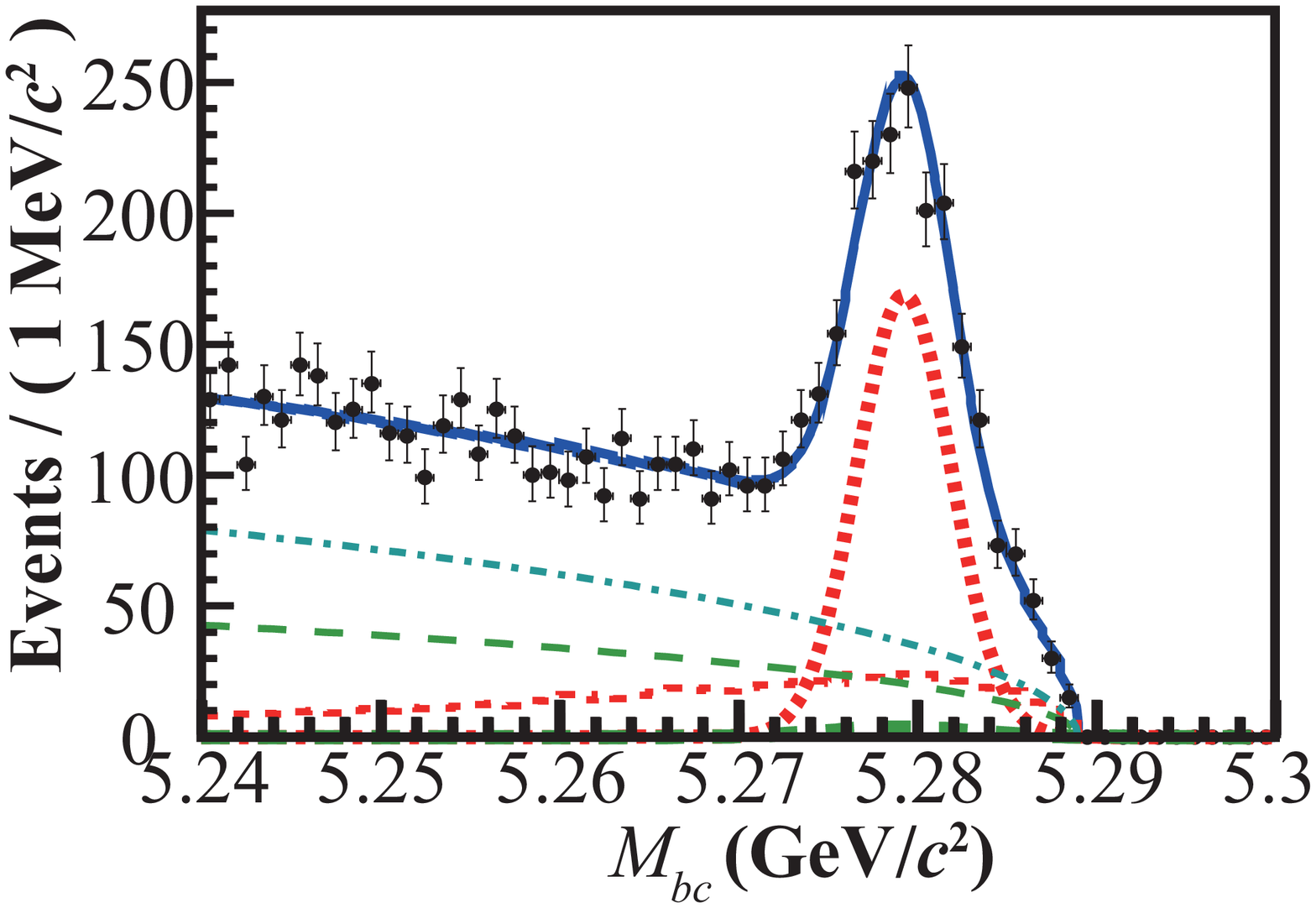} \label{fig:Mbcfit_MXs1.3-1.4}}
		\caption[$M_{\rm bc}$ fits in $M_{X_s}$ bins in 0.6\,GeV/$c^2\,<M_{X_s}<$\,1.4\,GeV/$c^2$. The plot shows the data (black points), and the fit function (blue solid line). The fit components correspond to signal (red thick short dashed line), cross-feed (red thin short dashed line), peaking {\BB} background (green thick long dashed line), non-peaking {\BB} background (green thin long dashed line) and {\qq} background (blue dot-dashed line).]
		{\small $M_{\rm bc}$ fits in $M_{X_s}$ bins in 0.6\,GeV/$c^2\,<M_{X_s}<$\,1.4\,GeV/$c^2$. The plot shows the data (black points), and the fit function (blue solid line). The fit components correspond to signal (red thick short dashed line), cross-feed (red thin short dashed line), peaking {\BB} background (green thick long dashed line), non-peaking {\BB} background (green thin long dashed line) and {\qq} background (blue dot-dashed line).}
		\label{fig:Mbcfit_MXs1}
	\end{center}	
\end{figure*}

\begin{figure*}[tb]
	\begin{center} 
	    \subfigure[1.4\,$<M_{X_s}<$\,1.5\,(GeV/$c^2$)] {\includegraphics[width=7.0cm]{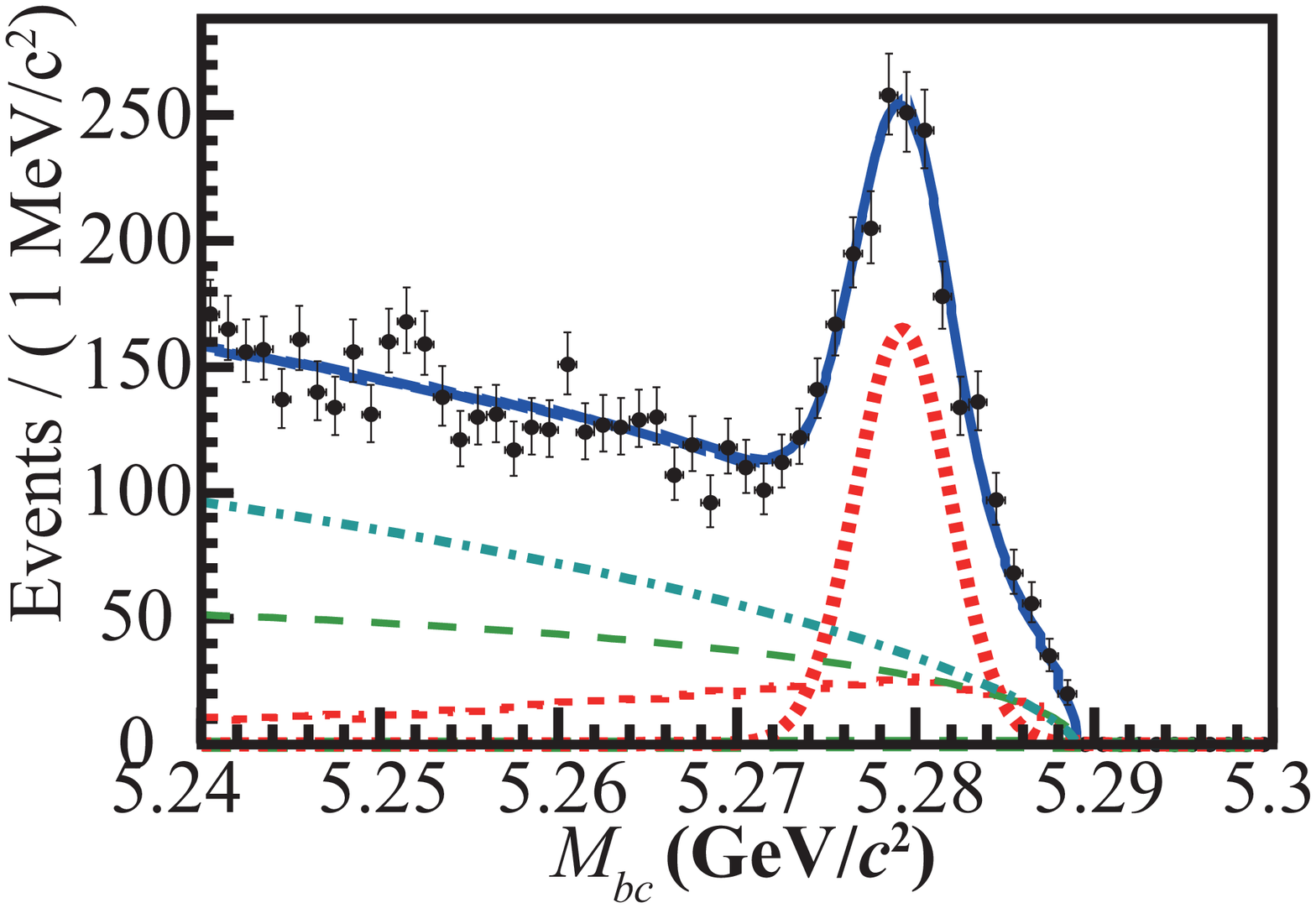} \label{fig:Mbcfit_MXs1.4-1.5}}
	    \subfigure[1.5\,$<M_{X_s}<$\,1.6\,(GeV/$c^2$)] {\includegraphics[width=7.0cm]{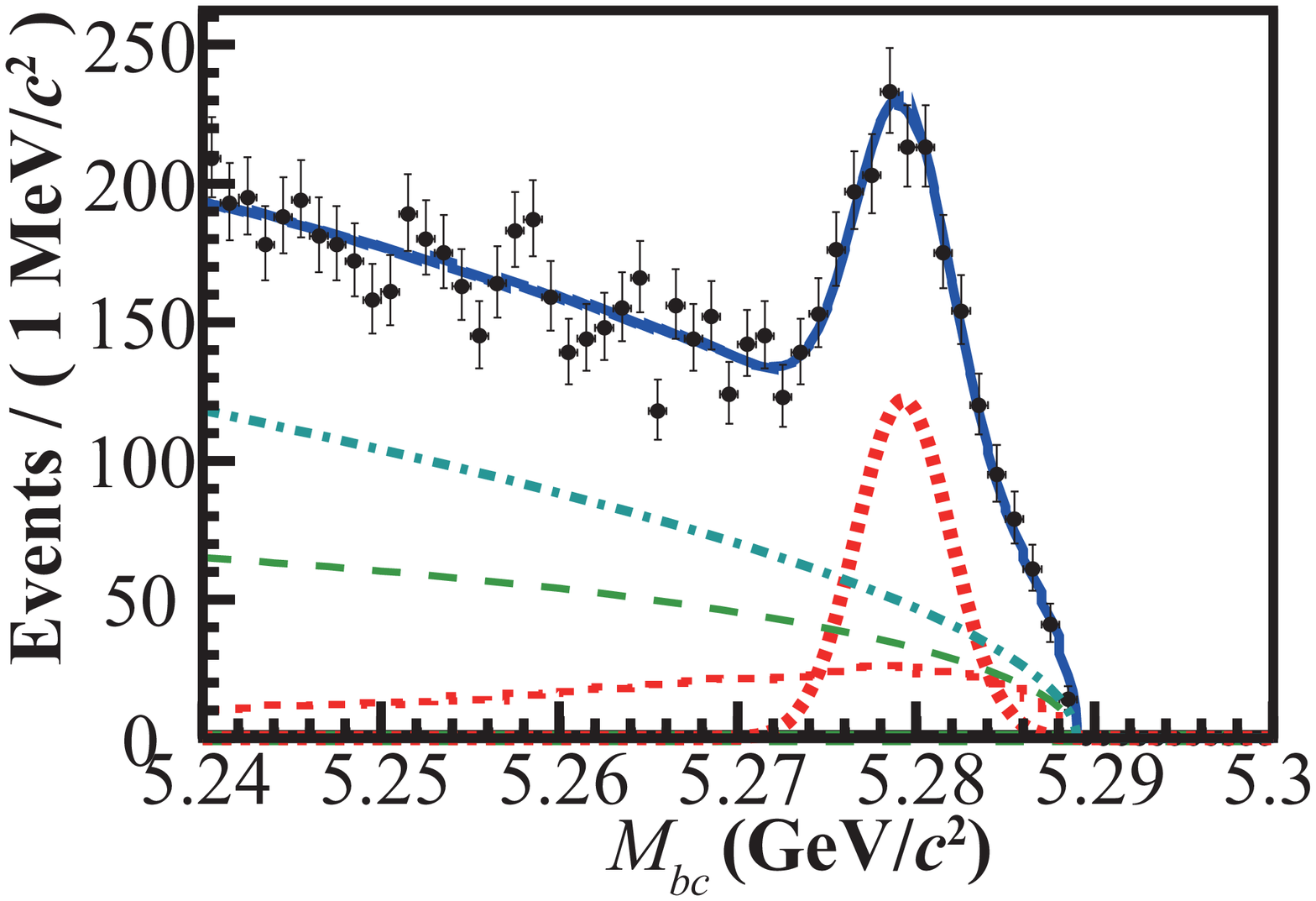} \label{fig:Mbcfit_MXs1.5-1.6}}
	    \subfigure[1.6\,$<M_{X_s}<$\,1.7\,(GeV/$c^2$)] {\includegraphics[width=7.0cm]{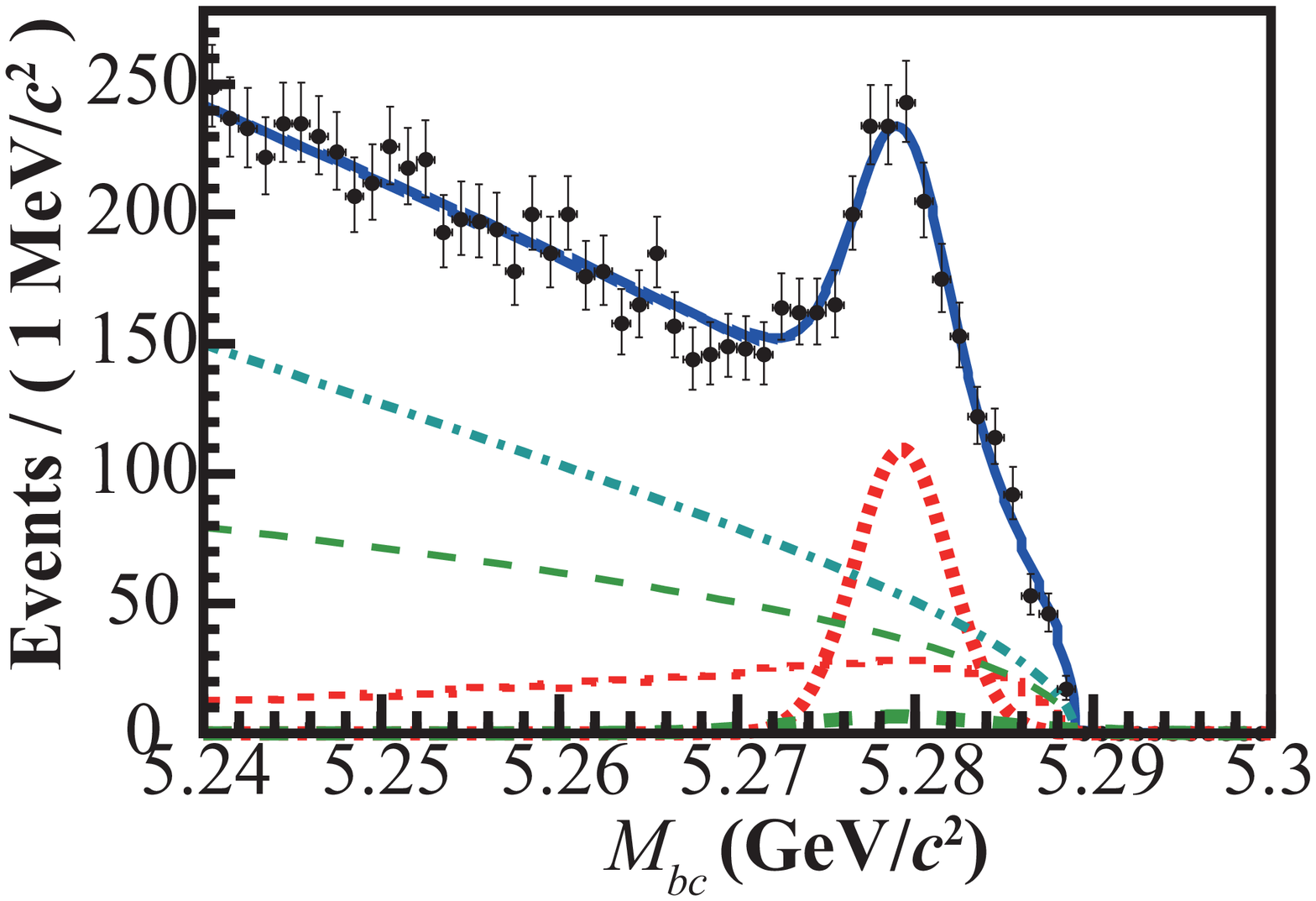} \label{fig:Mbcfit_MXs1.6-1.7}}
	    \subfigure[1.7\,$<M_{X_s}<$\,1.8\,(GeV/$c^2$)] {\includegraphics[width=7.0cm]{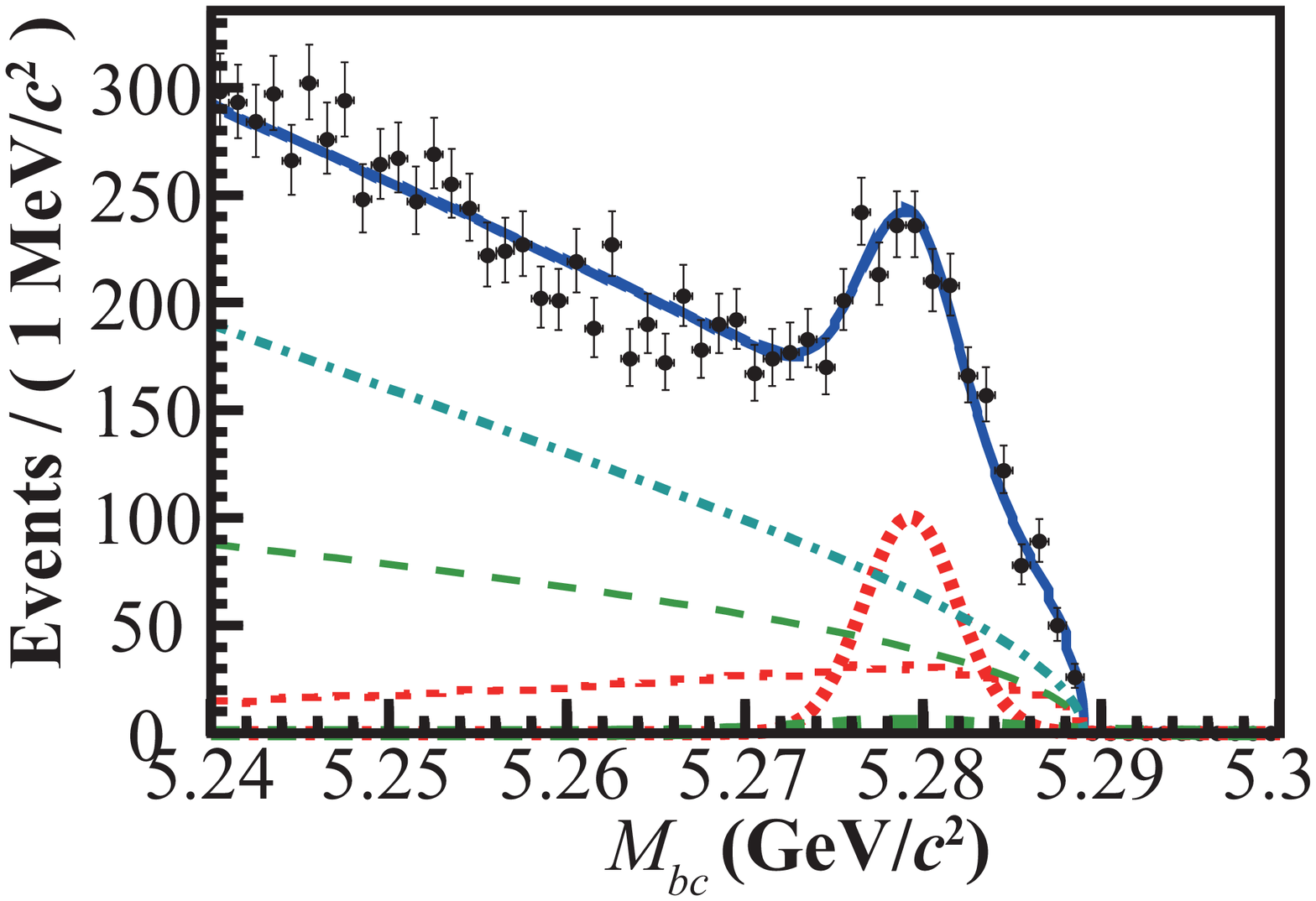} \label{fig:Mbcfit_MXs1.7-1.8}}
	    \subfigure[1.8\,$<M_{X_s}<$\,1.9\,(GeV/$c^2$)] {\includegraphics[width=7.0cm]{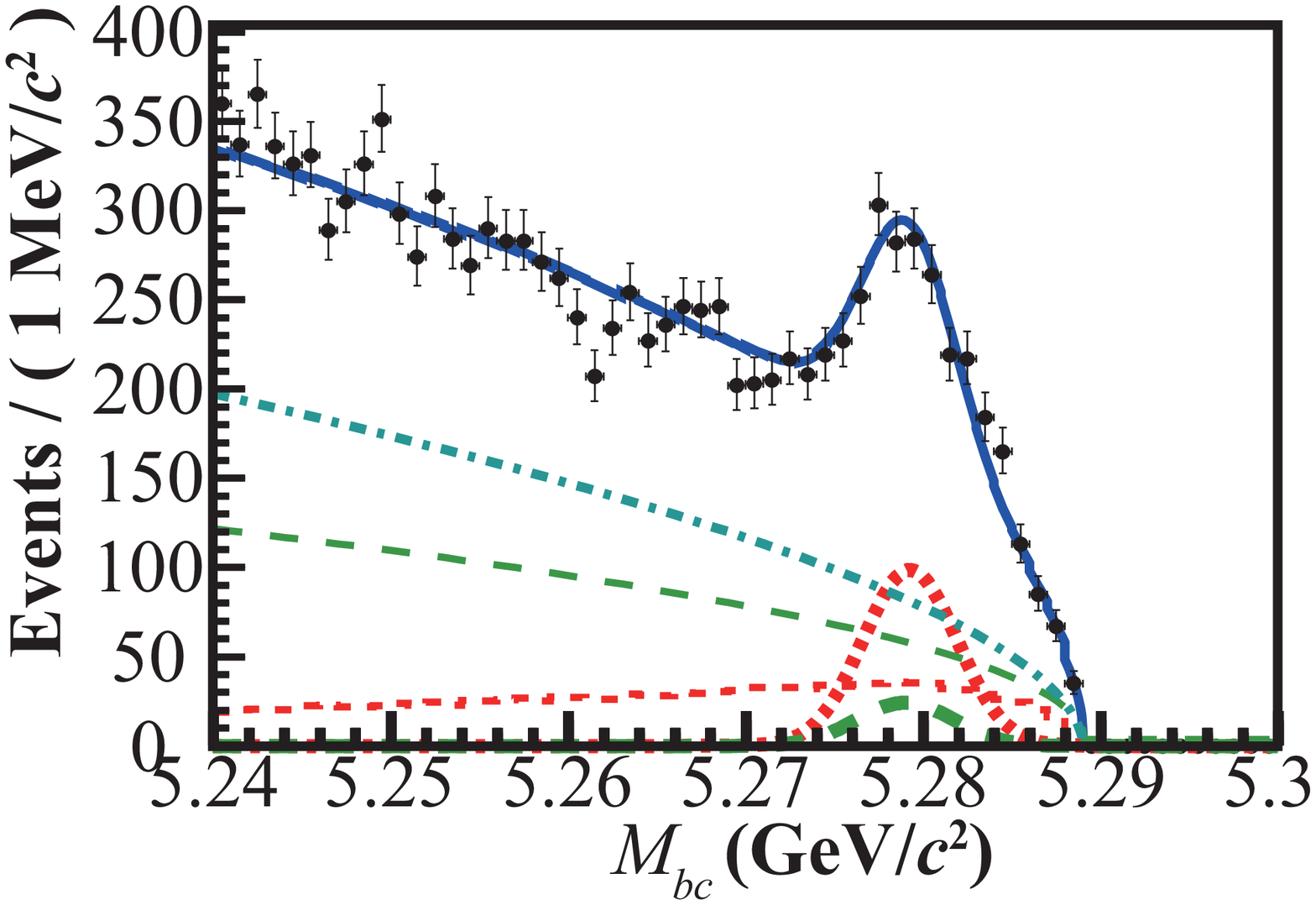} \label{fig:Mbcfit_MXs1.8-1.9}}
	    \subfigure[1.9\,$<M_{X_s}<$\,2.0\,(GeV/$c^2$)] {\includegraphics[width=7.0cm]{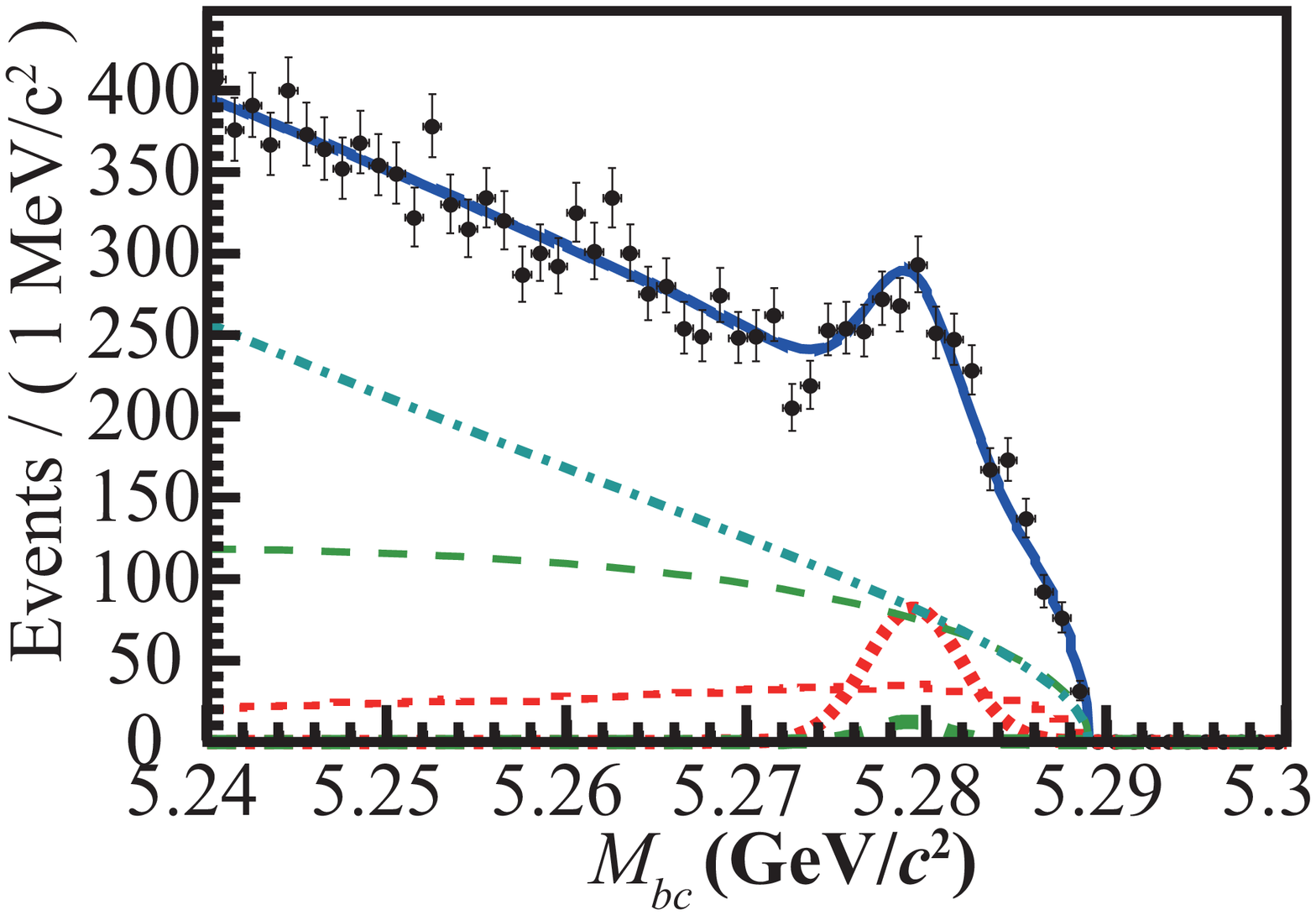} \label{fig:Mbcfit_MXs1.9-2.0}}
	    \subfigure[2.0\,$<M_{X_s}<$\,2.1\,(GeV/$c^2$)] {\includegraphics[width=7.0cm]{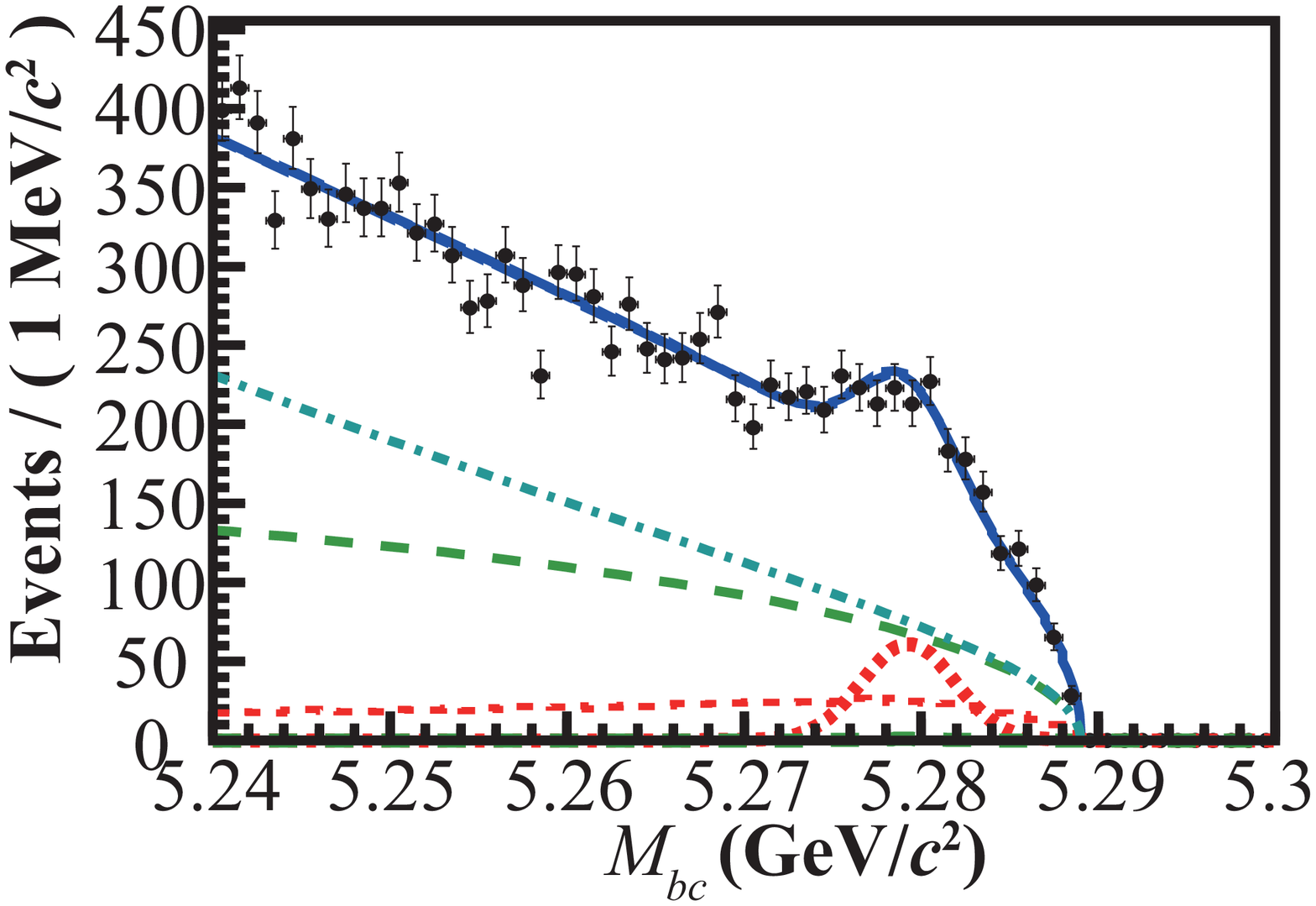} \label{fig:Mbcfit_MXs2.0-2.1}}
	    \subfigure[2.1\,$<M_{X_s}<$2.2\,(GeV/$c^2$)] {\includegraphics[width=7.0cm]{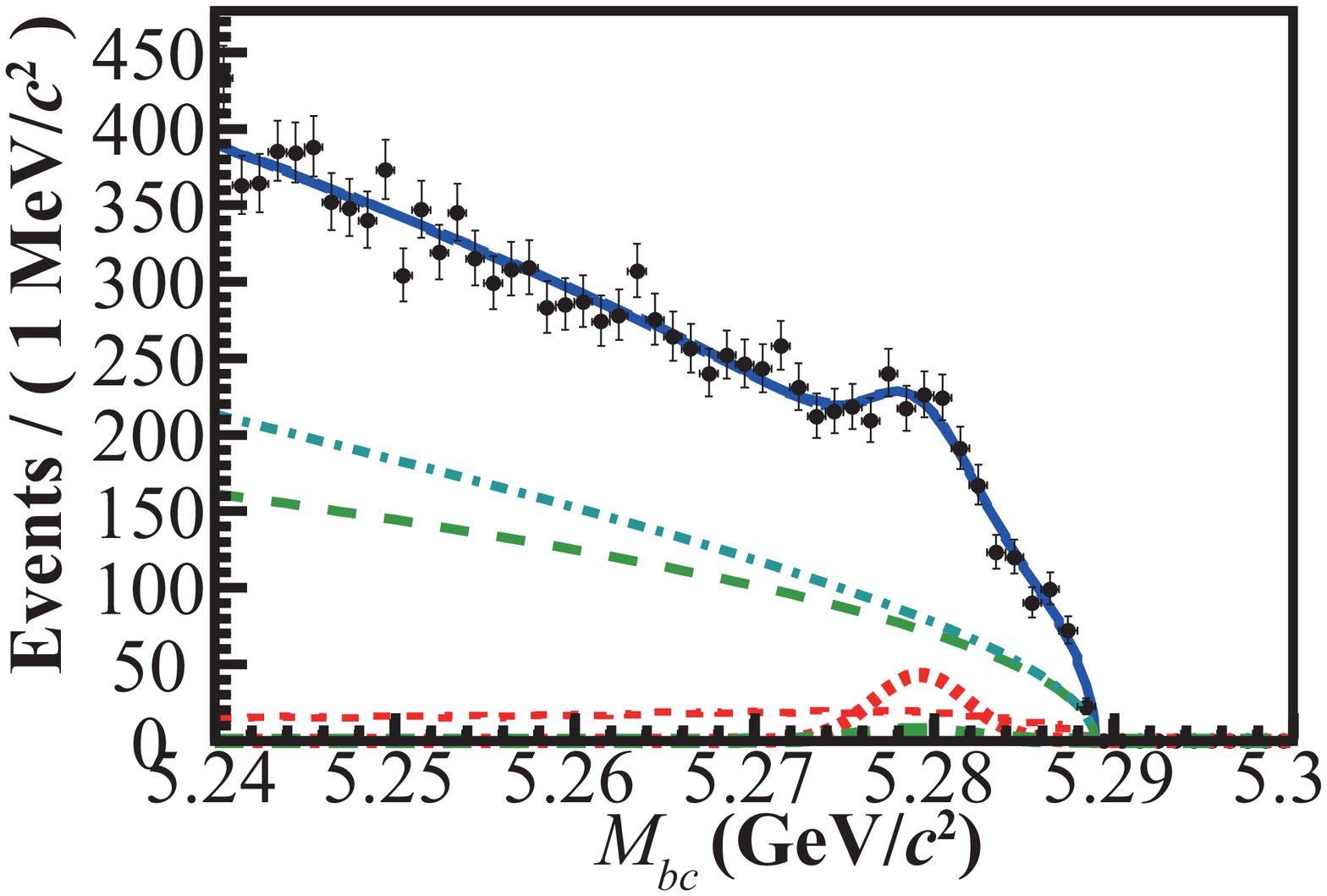} \label{fig:Mbcfit_MXs2.1-2.2}}
		\caption[$M_{\rm bc}$ fits in $M_{X_s}$ bins in 1.4\,GeV/$c^2\,<M_{X_s}<$\,2.2\,GeV/$c^2$. The plot shows the data (black points), and the fit function (blue solid line). The fit components correspond to signal (red thick short dashed line), cross-feed (red thin short dashed line), peaking {\BB} background (green thick long dashed line), non-peaking {\BB} background (green thin long dashed line) and {\qq} background (blue dot-dashed line).]
		{\small $M_{\rm bc}$ fits in $M_{X_s}$ bins in 1.4\,GeV/$c^2\,<M_{X_s}<$\,2.2\,GeV/$c^2$. The plot shows the data (black points), and the fit function (blue solid line). The fit components correspond to signal (red thick short dashed line), cross-feed (red thin short dashed line), peaking {\BB} background (green thick long dashed line), non-peaking {\BB} background (green thin long dashed line) and {\qq} background (blue dot-dashed line).}
        \label{fig:Mbcfit_MXs2}
	\end{center}	
\end{figure*}

\begin{figure*}[tb]
	\begin{center} 
	    \subfigure[2.2\,$<M_{X_s}<$\,2.4\,(GeV/$c^2$)] {\includegraphics[width=7.0cm]{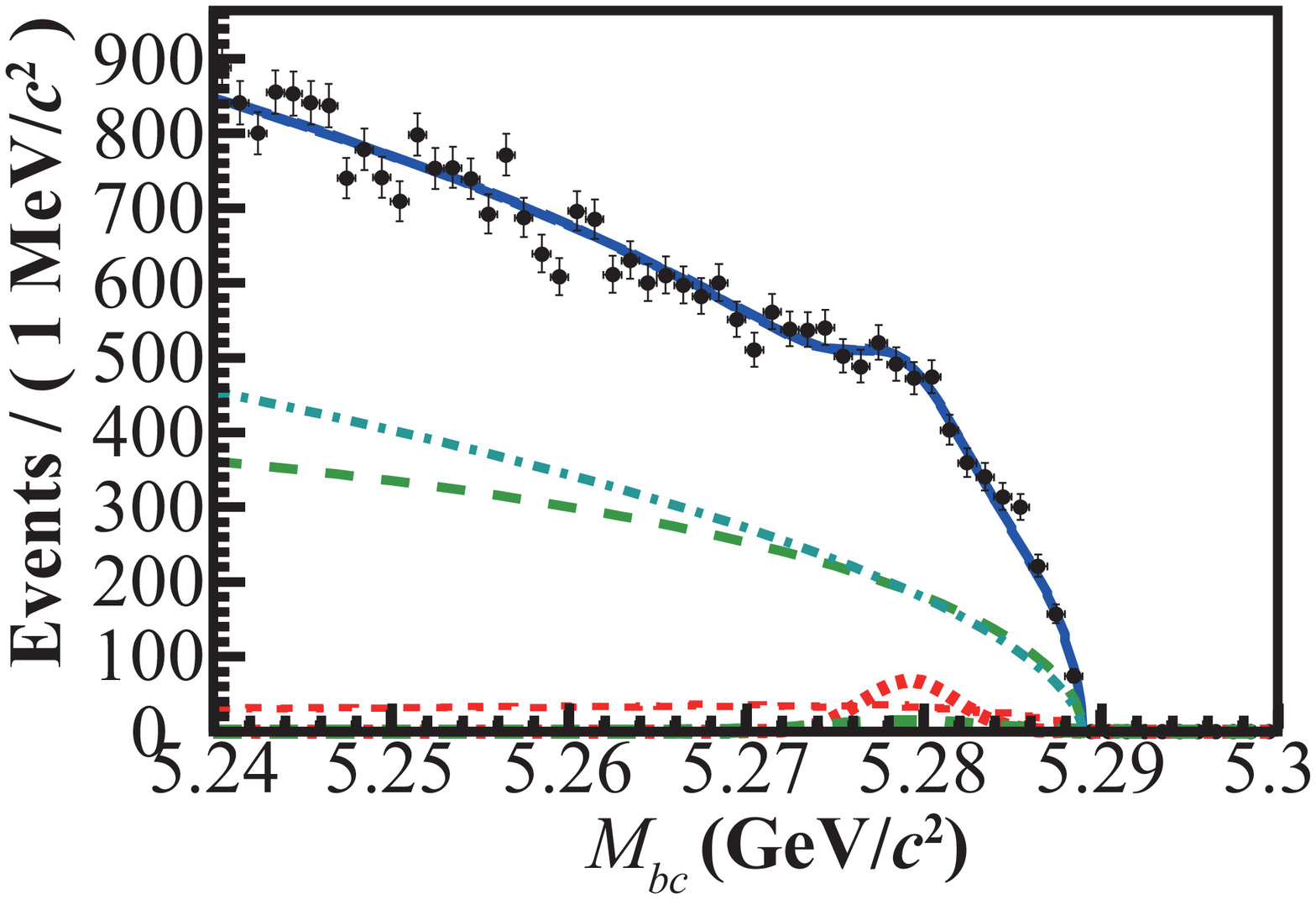} \label{fig:Mbcfit_MXs2.2-2.3}}
	    \subfigure[2.4\,$<M_{X_s}<$\,2.6\,(GeV/$c^2$)] {\includegraphics[width=7.0cm]{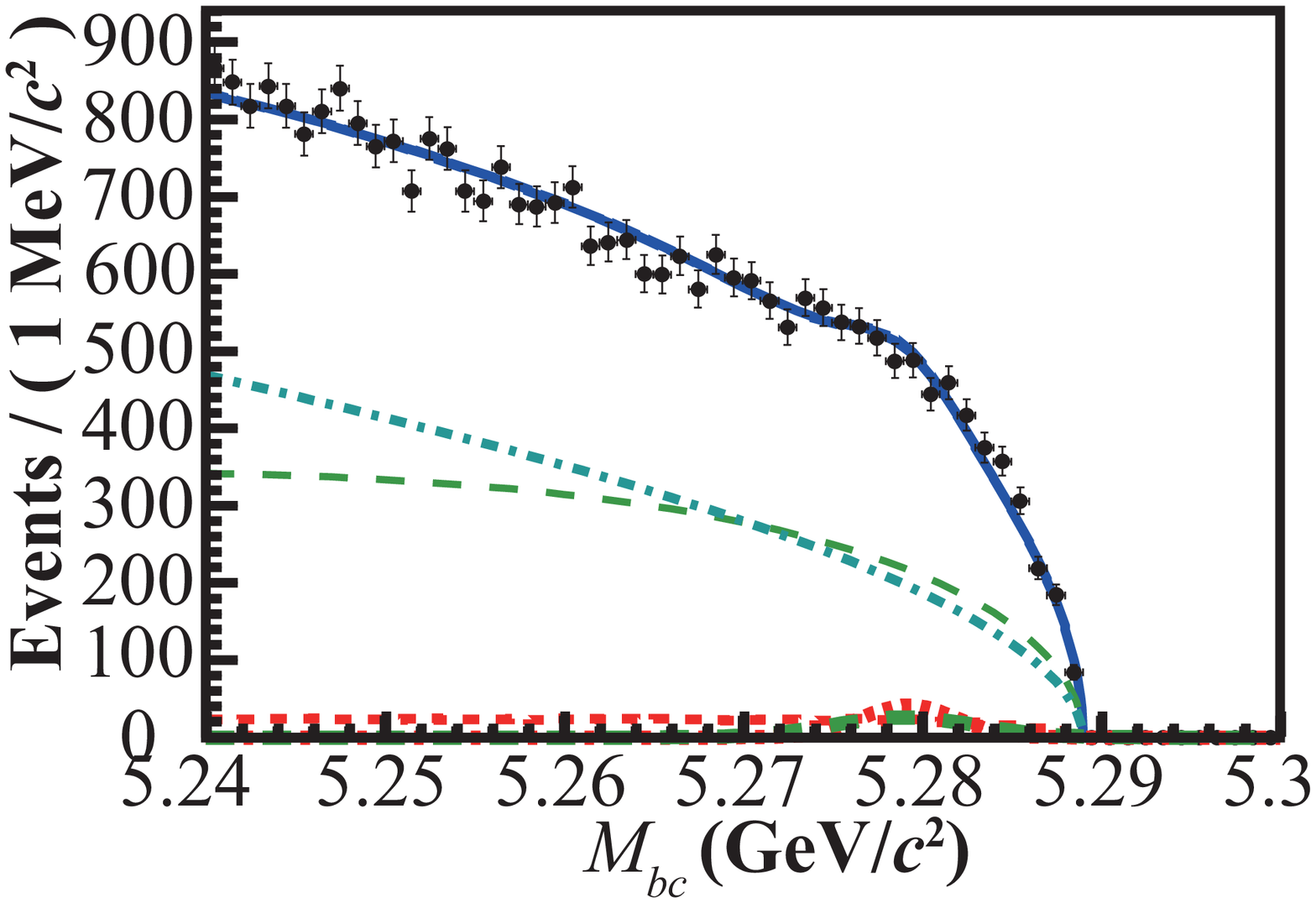} \label{fig:Mbcfit_MXs2.4-2.6}}
	    \subfigure[2.6\,$<M_{X_s}<$\,2.8\,(GeV/$c^2$)] {\includegraphics[width=7.0cm]{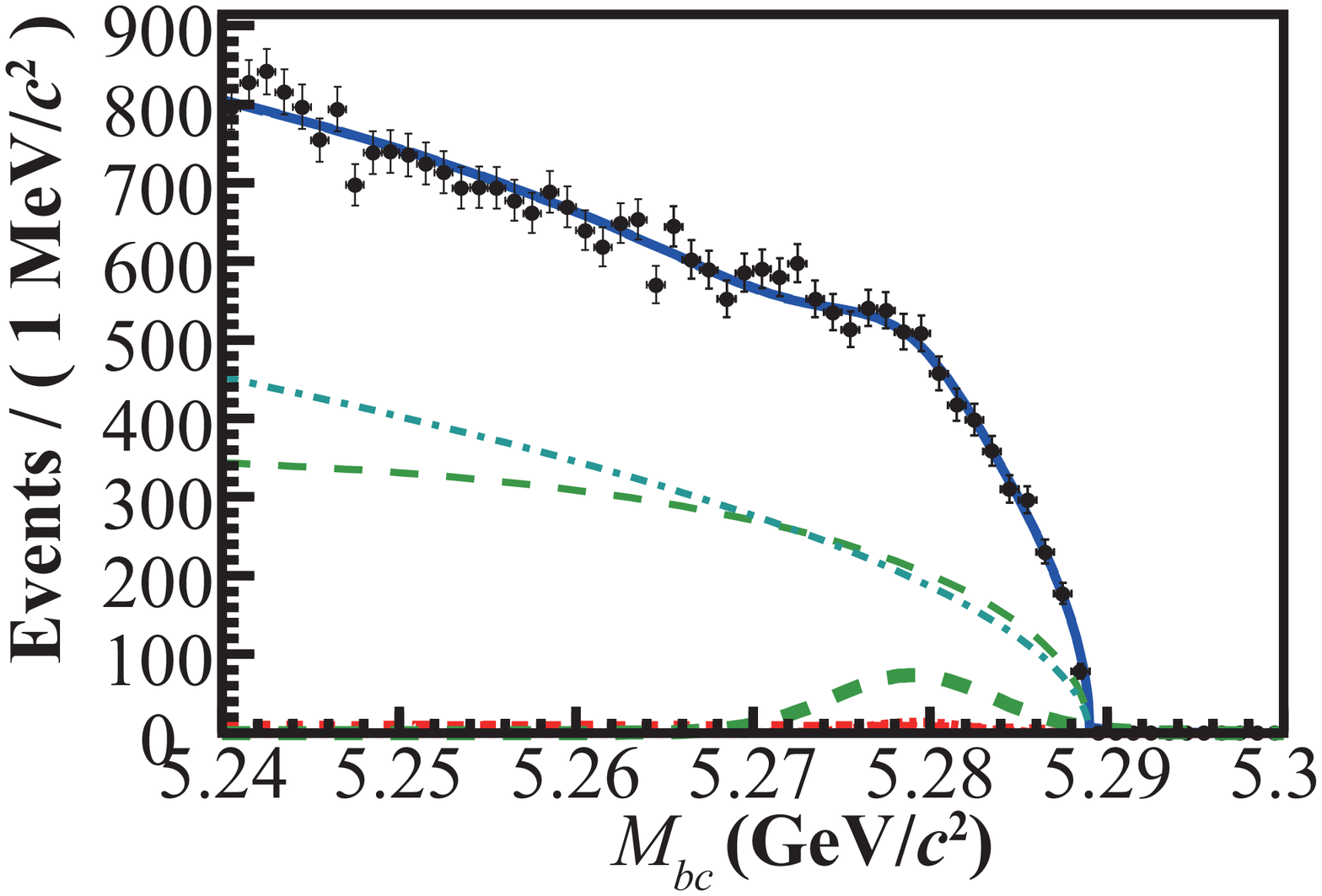} \label{fig:Mbcfit_MXs2.6-2.8}}
		\caption[$M_{\rm bc}$ fits in $M_{X_s}$ bins in 2.2\,GeV/$c^2\,<M_{X_s}<$\,2.8\,GeV/$c^2$. The plot shows the data (black points), and the fit function (blue solid line). The fit components correspond to signal (red thick short dashed line), cross-feed (red thin short dashed line), peaking {\BB} background (green thick long dashed line), non-peaking {\BB} background (green thin long dashed line) and {\qq} background (blue dot-dashed line).]
		{\small $M_{\rm bc}$ fits in $M_{X_s}$ bins in 2.2\,GeV/$c^2\,<M_{X_s}<$\,2.8\,GeV/$c^2$. The plot shows the data (black points), and the fit function (blue solid line). The fit components correspond to signal (red thick short dashed line), cross-feed (red thin short dashed line), peaking {\BB} background (green thick long dashed line), non-peaking {\BB} background (green thin long dashed line) and {\qq} background (blue dot-dashed line).}
		\label{fig:Mbcfit_MXs3}
	\end{center}	
\end{figure*}

\begin{figure}[tb]
	\begin{center}
		\includegraphics[width=8cm]{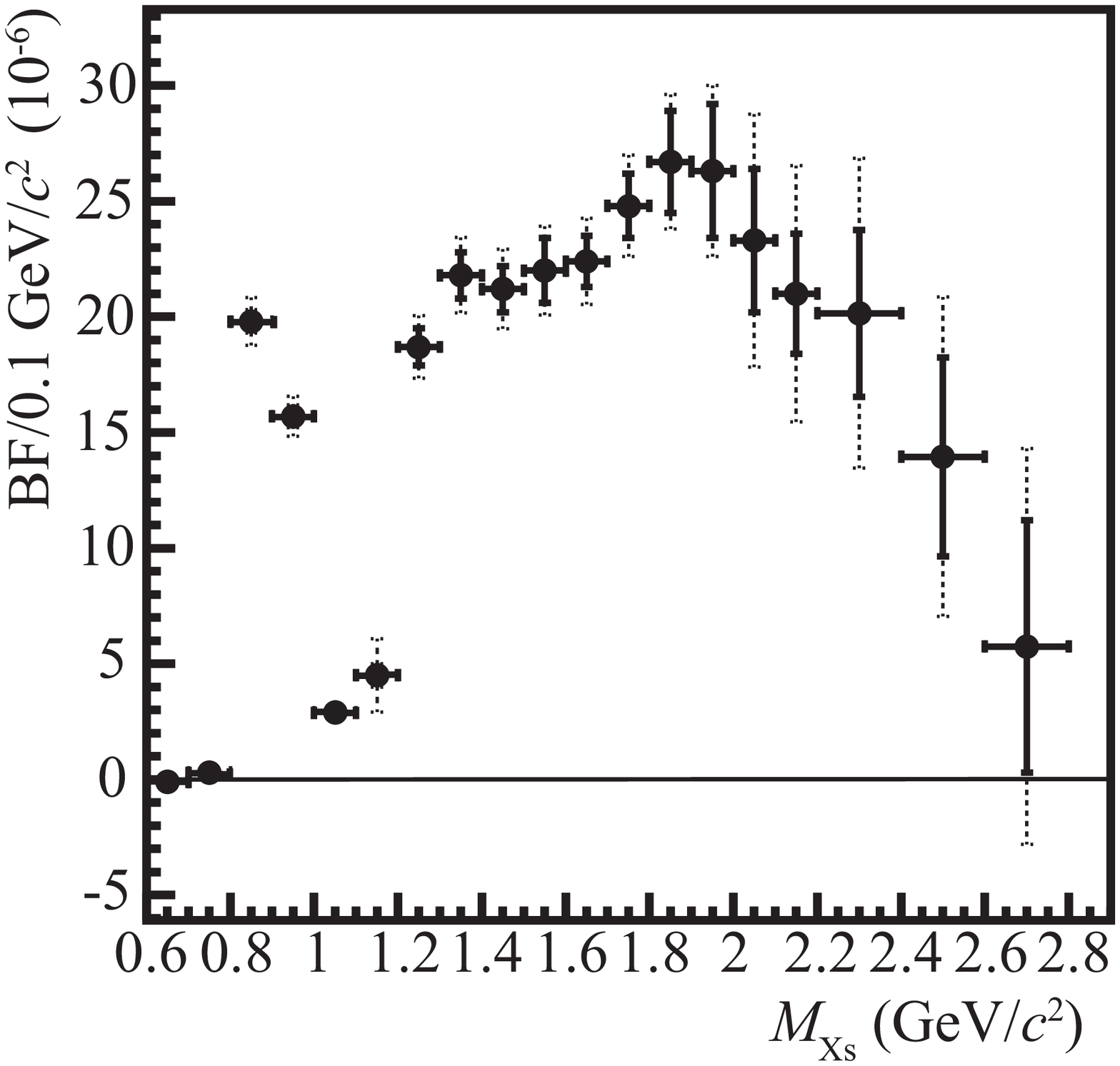}
		\caption[Partial branching fraction as a function of $M_{X_s}$. The error bars correspond to the statistical (solid) and the quadratic sum of the statistical and systematic (dashed).]
		{\small Partial branching fraction as a function of $M_{X_s}$. The error bars correspond to the statistical (solid) and the quadratic sum of the statistical and systematic (dashed).}
		\label{fig:PBF}
	\end{center}	
\end{figure}

The signal yields are obtained in 100\,MeV/$c^2$-wide bins in the low mass region, 0.6\,GeV/$c^2 <M_{X_s}<2.2$\,GeV/$c^2$, and 200\,MeV/$c^2$-wide bins in the high mass region, 2.2\,GeV/$c^2 < M_{X_s} <$\,2.8\,GeV/$c^2$.
The binned approach minimizes the sensitivity to modeling of the $X_s$ mass distribution.
In Table \ref{tb:BR_MXs}, the signal yields in each mass bin are reported from the fit to data.
Figures~\ref{fig:Mbcfit_MXs1}, \ref{fig:Mbcfit_MXs2} and \ref{fig:Mbcfit_MXs3} show the $M_{\rm bc}$ distribution fits in each $M_{X_s}$ bin. 
The partial branching fraction for each bin is defined as
\begin{eqnarray}
	{\cal B}_i = N_i / (N_{B\overline{B}} \times \epsilon_i) ,
\end{eqnarray}
where $N_i$ and $\epsilon_i$ are the signal yield and the efficiency, respectively, in bin $i$ and $N_{B\overline{B}}$ is the sum of number of $\overline{B^0}$ and $B^-$ events. The results are listed in Table \ref{tb:BR_MXs} and plotted in Fig.~\ref{fig:PBF}. 
Both statistical and systematic uncertainties in $M_{X_s}$ above 2.2\,GeV/$c^2$ are large. 
We also report the total branching fraction for $M_{X_s} <$\,2.8\,GeV/$c^2$, 
\begin{eqnarray}
	{\cal B}(\overline{B} \rightarrow X_s\gamma)=(3.51\pm0.17\pm0.33)\times10^{-4},
\end{eqnarray}
where the first uncertainty is statistical and the second is systematic. 
This branching fraction is the sum of 19 bins, with a 1.9 GeV lower threshold on the photon energy.
The total statistical uncertainty is based on the quadratic sum of the statistical uncertainty in each $X_s$ mass bin.
\begin{table}[tb]
	\begin{center}
		  \caption[The yields and partial branching fraction in each $M_{X_s}$ mass bin.]
		  {\small The yields and partial branching fraction in each $M_{X_s}$ mass bin.}
		  \label{tb:BR_MXs}
          \begin{tabular}{crr} \hline \hline
              $M_{X_s}$ bin &       Yield &${\cal B}\,(10^{-6})$  \\ 
              (GeV/$c^2$)   &             &         \\ \hline               
              0.6-0.7       & $-$6 $\pm$ 10 &   $-$0.1 $\pm$ 0.1 $\pm$ 0.0    \\ 
              0.7-0.8       &   36 $\pm$ 14 &      0.3 $\pm$ 0.1 $\pm$ 0.1    \\ 
              0.8-0.9       & 2032 $\pm$ 54 &     19.8 $\pm$ 0.5 $\pm$ 0.9    \\ 
              0.9-1.0       & 1689 $\pm$ 49 &     15.7 $\pm$ 0.5 $\pm$ 0.7    \\ 
              1.0-1.1       &  301 $\pm$ 27 &      2.9 $\pm$ 0.3 $\pm$ 0.2    \\ 
              1.1-1.2       &  310 $\pm$ 31 &      4.8 $\pm$ 0.5 $\pm$ 1.5    \\ 
              1.2-1.3       & 1019 $\pm$ 46 &     18.7 $\pm$ 0.8 $\pm$ 1.1    \\ 
              1.3-1.4       & 1117 $\pm$ 50 &     21.8 $\pm$ 1.0 $\pm$ 1.3    \\ 
              1.4-1.5       & 1090 $\pm$ 52 &     21.2 $\pm$ 1.0 $\pm$ 1.4    \\ 
              1.5-1.6       &  806 $\pm$ 50 &     22.0 $\pm$ 1.4 $\pm$ 1.3    \\ 
              1.6-1.7       &  723 $\pm$ 37 &     22.4 $\pm$ 1.1 $\pm$ 1.5    \\ 
              1.7-1.8       &  664 $\pm$ 37 &     24.8 $\pm$ 1.4 $\pm$ 1.7    \\  
              1.8-1.9       &  652 $\pm$ 54 &     26.7 $\pm$ 2.2 $\pm$ 1.9    \\ 
              1.9-2.0       &  542 $\pm$ 60 &     26.3 $\pm$ 2.9 $\pm$ 2.3    \\  
              2.0-2.1       &  403 $\pm$ 54 &     23.3 $\pm$ 3.1 $\pm$ 4.5    \\ 
              2.1-2.2       &  285 $\pm$ 35 &     21.0 $\pm$ 2.6 $\pm$ 4.9    \\   
              2.2-2.4       &  449 $\pm$ 80 &     40.3 $\pm$ 7.2 $\pm$ 11     \\ 
              2.4-2.6       &  273 $\pm$ 84 &     27.9 $\pm$ 8.6 $\pm$ 11     \\ 
              2.6-2.8       &   87 $\pm$ 82 &     11.5 $\pm$ 11 $\pm$ 13     \\ \hline \hline           
          \end{tabular}
	\end{center}
\end{table}

\section{Conclusion}
We measure the branching fraction of {\BXsg} with the sum-of-exclusives approach using the entire {\upshiron} data set of the Belle experiment. 
The branching fraction in the region $M_{X_s}<$\,2.8\,GeV/$c^2$ (corresponding to a minimum photon energy of 1.9 GeV) is measured to be	${\cal B}(\overline{B} \rightarrow X_s\gamma)=(3.51\pm0.17\pm0.33)\times10^{-4}$, where the first uncertainty is statistical and the second is systematic. 
This result is consistent with the measurement at BABAR \cite{Babar_semiincl} and achieves the best precision of any sum-of-exclusives approach.
This measurement supersedes our previous result \cite{ushiroda}. 

\section{Acknowledgments}
We thank the KEKB group for the excellent operation of the
accelerator; the KEK cryogenics group for the efficient
operation of the solenoid; and the KEK computer group,
the National Institute of Informatics, and the 
PNNL/EMSL computing group for valuable computing
and SINET4 network support.  We acknowledge support from
the Ministry of Education, Culture, Sports, Science, and
Technology (MEXT) of Japan, the Japan Society for the 
Promotion of Science (JSPS), and the Tau-Lepton Physics 
Research Center of Nagoya University; 
the Australian Research Council;
Austrian Science Fund under Grant No.~P 22742-N16 and P 26794-N20;
the National Natural Science Foundation of China under Contracts 
No.~10575109, No.~10775142, No.~10825524, No.~10875115, No.~10935008 
and No.~11175187; 
the Ministry of Education, Youth and Sports of the Czech
Republic under Contract No.~LG14034;
the Carl Zeiss Foundation, the Deutsche Forschungsgemeinschaft
and the VolkswagenStiftung;
the Department of Science and Technology of India; 
the Istituto Nazionale di Fisica Nucleare of Italy; 
National Research Foundation of Korea Grants
No.~2011-0029457, No.~2012-0008143, No.~2012R1A1A2008330, 
No.~2013R1A1A3007772, No.~2014R1A2A2A01005286, No.~2014R1A2A2A01002734, 
No.~2014R1A1A2006456;
the BRL program under NRF Grant No.~KRF-2011-0020333, No.~KRF-2011-0021196,
Center for Korean J-PARC Users, No.~NRF-2013K1A3A7A06056592; the BK21
Plus program and the GSDC of the Korea Institute of Science and
Technology Information;
the Polish Ministry of Science and Higher Education and 
the National Science Center;
the Ministry of Education and Science of the Russian
Federation and the Russian Federal Agency for Atomic Energy;
the Slovenian Research Agency;
the Basque Foundation for Science (IKERBASQUE) and the UPV/EHU under 
program UFI 11/55;
the Swiss National Science Foundation; the National Science Council
and the Ministry of Education of Taiwan; and the U.S.\
Department of Energy and the National Science Foundation.
This work is supported by a Grant-in-Aid from MEXT for 
Science Research in a Priority Area (``New Development of 
Flavor Physics'') and from JSPS for Creative Scientific 
Research (``Evolution of Tau-lepton Physics'').

\end{document}